\documentclass[twocolumn,nobibnotes,aps,pre,showpacs]{revtex4-2}
\usepackage{hyperref}
\usepackage{graphics}
\usepackage{graphicx}
\usepackage{color}
\usepackage{amsmath}
\usepackage{amssymb}
\usepackage{bm}
\usepackage{float}
\usepackage{epsfig}
\usepackage{epstopdf}
\usepackage{color,graphics}
\usepackage{calrsfs}
\usepackage{bbold}
\usepackage[bottom]{footmisc}

\newcommand\be{\begin{equation}}
\newcommand\e{\end{equation}}
\newcommand\ba{\begin{eqnarray}}
\newcommand\ay{\end{eqnarray}}

\begin{document}

\title{Methodological notes on the gauge invariance in the treatment of waves and oscillations in plasmas $via$ the Einstein-Vlasov-Maxwell system: Fundamental equations}
\author{Lucas Bourscheidt}
\affiliation{Physics Institute, Federal University of Rio Grande do Sul, CEP 91501-970,
Av. Bento Gon\c{c}alves 9500, Porto Alegre, RS, Brazil}
\author{Fernando Haas}
\affiliation{Physics Institute, Federal University of Rio Grande do Sul, CEP 91501-970,
Av. Bento Gon\c{c}alves 9500, Porto Alegre, RS, Brazil}

\begin{abstract}
The theory of gauge transformations in linearized gravitation is investigated. After a brief discussion of the fundamentals of the kinetic theory in curved spacetime, the Einstein-Vlasov-Maxwell (EVM) system of equations in terms of gauge invariant quantities is established without neglecting the equations of motion associated with the dynamics of the non-radiative components of the metric tensor. The established theory is applied to a non-collisional electron-positron plasma, leading to a dispersion relation for gravitational waves in this model system. The problem of Landau damping is addressed and some attention is given to the issue of the energy exchanges between the plasma and the gravitational wave. In a future paper, a more complete set of approximate dispersion relations for waves and oscillations in plasmas will be presented, including the dynamics of non-radiative components of the metric tensor, with special attention to the problems of the Landau damping and of the energy exchanges between matter, the electromagnetic field and the gravitational field.
\end{abstract}

\maketitle

\section{Introduction}

The questions about the propagation and even the existence of gravitational waves go back to the very foundations of the general theory of relativity \cite{eins1,eins2}. These waves have been well investigated (both experimentally and theoretically \cite{sch1,sch2,taylor1,taylor2,weinberg,mis,web,ciuf,pra,bart,barish,coop,madore,grif}) over the past decades, culminating in their detection some years ago by the LIGO system \cite{ligo1,ligo2,nsmerger,soares}. From a theoretical point of view, the discussion of gravitational waves is very facilitated and simplified in the linear regime \cite{weinberg,mis,web,ciuf}. In this limiting case, it is relatively simple to show that gravitation behaves most like electromagnetism, in the sense that Einstein equations exhibits gauge freedom, in the same way as Maxwell equations, and that the gravitational waves possesses, as electromagnetic waves do, two independent states of polarization. On the other hand, it is well established in the realm of the classic field theory that only gauge invariant quantities could have a physical significance \cite{jack,sche}, and one of the objectives of this manuscript is to exploit this subject and properly apply it to the study of waves and oscillations in fully relativistic plasmas in the context of the kinetic theory \cite{cerci}. 

The history of the theoretical study of the propagation of gravitational waves in plasmas began long before the experimental confirmation of their existence. In the 1970s, the problem of gravitational wave propagation in a medium was addressed by considering a hot non-collisional system of particles through kinetic theory \cite{chesters}. In this study, a dispersion relation was derived by the resolution of the Einstein-Boltzmann (in its non-collisional version, that is, Vlasov equation) system and it was concluded, on the one hand, that the impact of the plasma on the propagation of gravitational waves should be small and, on the other hand, that there is a possibility of wave-particle resonances in this physical system. In later works, the controversy over the possibility of Landau damping of gravitational waves was discussed at several levels \cite{gay}, and the effect of gravitational radiation from distant sources on electromagnetic waves in dispersive media was investigated \cite{bert}. A complete system of macroscopic equations for the electromagnetic and gravitational fields in magnetized plasmas was presented in 1982 and the propagation of gravitational waves in this medium was investigated using kinetic theory in the following year \cite{bre,mac}. In the late 1990s, a system of equations governing the nonlinear dynamics of interacting neutrinos and gravitons in plasmas was established \cite{mark1} and the excitation of electromagnetic and Langmuir waves by gravitational waves was considered \cite{mark2}.

At the turn of the century, the effects of intense gravitational waves (ie, in the nonlinear regime) on cold plasmas have been investigated \cite{mark3}. Then, in a series of works, the interactions between gravitational waves, electromagnetic waves and a magnetized plasma were explored, and the issue of gravitational Landau damping was raised again \cite{mark4,mark5,serv1,moort1,mof}. In the same period, some studies were carried out to examine the excitation of magnetosonic waves and the transfer of energy from gravitational waves to the plasma, especially in neutron star merger events \cite{papa1,moort2,vla1,vla2}. Furthermore, cosmological issues involving the excitation of plasma waves by gravitational waves have been considered in 2002, employing relativistic hydrodynamic equations \cite{papa2}. In 2004, also with a hydrodynamic approach, the nonlinear coupling between Alfvén waves and gravitational waves in strongly magnetized plasmas was scrutinized \cite{kall}. Spherically symmetric solutions of the Einstein-Vlasov-Maxwell system have been discussed in 2004-05 \cite{nond1,nond2}, and the transfer of gravitational energy to plasma particles in a system of astrophysical interest and the interaction between gravitational waves in strongly magnetized plasmas were topics revisited in 2006 \cite{vla3,isl}. In 2010, in a sequence of works, coupling constants for the nonlinear interactions between gravitational, electrostatic and electromagnetic waves in a relativistic non-magnetized and non-collisional plasma were obtained, and a system of coupled equations for gravitational and electromagnetic waves in a relativistic plasma have been analyzed allowing the identification of several resonances between gravitational waves and plasma \cite{mark6,fos}.

The effects of dispersion and scattering through the interstellar medium on the detection of low-frequency gravitational waves have been analyzed in 2013 \cite{stin}. In 2015, vorticity generation in a plasma around Schwarzschild and Kerr black holes was investigated employing a magnetofluid and ADM (Arnowitt-Deser-Misner) formalisms \cite{adm}. In this work, a prescribed metric was assumed and some results of astrophysical interest were obtained, such as a proposed mechanisms for collimation of plasma jets and for vortex formation in the protoplanetary disks of a supermassive star, presumably related to planet formation \cite{bhatt1}. In 2017, the damping of gravitational waves in collisional material media was investigated through the Boltzmann equation, evidencing two distinct mechanisms for the process \cite{baym}. In the same year, the coupling between the electromagnetic and gravitational fields was once again the subject of study, possibly motivated by (at the time) recent observations of the LIGO-Virgo collaboration \cite{cabr,dol}. In 2018, the magnetofluid formalism was revisited in a work where plasmas was taken as a system of multiple perfect charged fluids, and the formalism proposed provided suitable tools for characterizing plasmas in a given curved space–time, including equilibrium states \cite{bhatt2}. No directly related to gravitational radiation (but still a very interesting subject), the energy extraction from a spinning Kerr black hole via magnetic reconection was theoretically investigated, showing that the process is possible for black holes of high spin surrounded by a strongly magnetized plasma \cite{comisso}. In recent years, some research topics in the area have been the excitation of magnetohydrodynamic waves by gravitational waves in strongly magnetized plasmas \cite{gont}, the dissipation of gravitational waves \cite{loeb} and the propagation of gravitational waves in magnetized dielectric media \cite{moro}. Very recently, some works have focused on the important issue of the polarization of gravitational waves \cite{and,kah}, and a mechanism for the reflection of electromagnetic waves by a gravitational wave background in plasma media was proposed \cite{ansejo}.

Some excellent theses developed over the last two decades and which certainly also deserve mention are the works of M. Servin \cite{serv2}, J. B. Moortgat \cite{moort3} and O. Janson \cite{jan1}. These texts all deal essentially with the same topic: the propagation and interactions of electromagnetic and gravitational waves in plasmas in the context of general relativity. As we have seen, in the present work we intend to answer questions about the behavior of this same class of physical system and, therefore, these three works are fundamental to us.

Our aim in this work is to bring together, in a clear, concise, direct and (as far as possible) self-contained way, the fundamentals of the classical theory of fields - specially the issue of the gauge freedom - and the most profound theory of the plasma state (in a non-quantum level) which, in our understanding, is the general relativistic kinetic theory. This approach allows the safe and unambiguous determination of the radiative components of the gravitational field, without neglecting the dynamics associated with the other (non-radiative) components. Furthermore, an approximate dispersion relation for gravitational waves in an electron-positron plasma, obtained from the EVM system of equations, will be treated and discussed in order to illustrate the established methodology and, to some extent, to elucidate the behavior of gravitational waves in plasmas. In short, our goal is to establish a complete picture of the dynamics and interactions of gravitation, electromagnetism, and matter in plasma state using only gauge invariant quantities whenever possible, and apply the formalism to find out dispersion relations for any kind of gravitational oscillation, including the calculation of their damping and energy exchange rates. 

We need to reinforce that the present observational and experimental status of plasma physics does not require a covariant formulation (since gravitational effects are generally small), except when the aim is to describe the dynamics of components of the metric tensor that do not have a Newtonian counterpart, as is the case of the present paper. In addition, our formalism can have application e.g. to cosmological plasmas involving small-scale dynamos in Riemannian spaces. This is the case of Lobachevsky or spherical geometries, which are possible geometries for the spatial part of the Friedmann cosmological models \cite{sokoloff}.

The manuscript is organized in the following way. In Section II, Einstein field equations (that is, the equations of gravitation) are established in the linear (weak field) regime, and the gauge freedom of the theory is discussed in some detail. The equations of motion are then written in terms of gauge invariants analogous to the electric and magnetic fields of the Maxwell theory, and the general theory of gravitational waves in terms of these invariants is discussed. Some attention is payed to classical field theoretical aspects of gravitation and to the energy carried and exchanged by the gravitational field. In Section III, we digress into the general relativistic kinetic theory of non-collisional plasmas and write down the Vlasov equation which, together with Einstein and Maxwell equations, constitute a complete system of equations for the description of plasma state phenomena. In that Section, the equations are written in terms of the aforementioned gauge invariants whenever possible. Finally, in Section IV, an dispersion relations for gravitational waves in an electron-positron plasma is obtained and discussed.

When writing this manuscript we keep in mind a wide public, specially physicist that, like us, are interested in both branches of physics (plasma physics and general relativity). The step-by-step construction shown in the text reflect very much the path taken by the authors in preparing these methodological notes, and are presented precisely because, presumably, other researchers in plasma physics interested in general relativity (especially in gravitational waves) but without much experience in the area can benefit from this. Therefore, it is important to emphasize that a reader specialized in general relativity can safely skip some parts of the text, especially Section II and part A of Section III.

We follow the convention in which the signature of the metric tensor is $(+ - - -)$ and the Ricci tensor $R_{\mu\nu}$ is obtained from contracting the first and the last indices of the Riemann-Christoffel tensor, ${R^\sigma}_{\mu\nu\rho}$. Latin indices can assume values from 1 to 3 and are used do label spatial coordinates. Greek indices can assume values from 0 to 3 and are used to label space-time coordinates. The temporal coordinate is denoted by $x^0 = ct$, where $c$ is the velocity of light and $t$ is the time. Einstein summation convention is assumed. It should be clear from the context when a superscript symbolizes an index or an exponent.

\section{Gauge transformations in linearized gravitation}

\subsection{On the weak field limit of the Einstein equations}

We proceed to a brief discussion of the equations of gravitation in the weak field limit \cite{weinberg,mis,web}. This treatment provides a linearized version of Einstein equations, facilitating – or even allowing – a general and detailed study of gravitational waves, the main object of study of this work. In particular, the Fourier decomposition of the field is legitimate only in the linearized theory. In addition to the non-linearity of Einstein equations, the theoretical study of gravitational waves is hampered by a subtlety whose origin lies in the very general covariance of the theory: as the choice of coordinate system is completely arbitrary, it can be difficult to distinguish which wave solutions represent real physical effects and which are mere artifices resulting from a particular choice of coordinates. This last difficulty will be properly addressed in part B of the present Section.

Usually, the full Einstein equations (neglecting the cosmological term) are written in the form
\begin{equation}
G_{\mu\nu} = -\frac{8\pi G}{c^4} T_{\mu\nu} \,\label{e1}
\end{equation}
with the Einstein tensor $G_{\mu\nu}$ given by
\begin{equation}
G_{\mu\nu} = R_{\mu\nu} -\frac{1}{2}g_{\mu\nu}R \,,\label{e2}
\end{equation}
where $R_{\mu\nu}$, $R$, $g_{\mu\nu}$, and $T_{\mu\nu}$ are, respectively, the Ricci curvature tensor, the Ricci curvature scalar, the metric tensor (whose components are the gravitational potentials), and the stress-energy tensor of the physical system, which constitute the source of the gravitational field. As usual, $c$ is the velocity of light in vacuum and $G$ is the Newtonian gravitational constant. As is well known, in general relativity gravitation is conceived properly as a manifestation of the curvature of space-time caused by a distribution of mass and energy, both contained in $T_{\mu\nu}$. Curvature itself is encapsulated primarily in the metric tensor $g_{\mu\nu}$ and secondly in the Ricci mathematical objects $R_{\mu\nu}$ and $R$ (among others). Under the exclusive influence of a gravitational field, a point particle of mass $m$ moves in space-time along geodesic trajectories, which are the curved space analogous of straight lines of the flat (or empty) space. The equations of motion of the point particle are in this case
\begin{equation}
\frac{d^2x^{\mu}}{d\tau^2} + \Gamma_{\nu\rho}^{\mu} \frac{dx^{\nu}}{d\tau} \frac{dx^{\rho}}{d\tau} = 0 \,,\label{e3}
\end{equation}
where the derivatives of the coordinates $x^{\mu}$ are taken with respect to the proper time $\tau$, and $\Gamma_{\nu\rho}^{\mu}$ are the Christoffel symbols of the second kind, given in terms of derivatives of the metric by
\begin{equation}
\Gamma_{\nu\rho}^{\mu} = \frac{1}{2}g^{\mu\sigma} \left (\frac{\partial g_{\nu\sigma}}{\partial x^{\rho}} + 
\frac{\partial g_{\rho\sigma}}{\partial x^{\nu}} -
\frac{\partial g_{\nu\rho}}{\partial x^{\sigma}}
\right ) \,.\label{e4}
\end{equation}
If, besides the gravitation, the particle is subjected to an electromagnetic field $F_{\mu\nu}$, the equations of motion must be modified to incorporate the electromagnetic force, taking the form
\begin{equation}
\frac{d^2x^{\mu}}{d\tau^2} + \Gamma_{\nu\rho}^{\mu} \frac{dx^{\nu}}{d\tau} \frac{dx^{\rho}}{d\tau} = \frac{q}{m} {F^\mu}_{\nu} \frac{dx^{\nu}}{d\tau} \,,\label{e5}
\end{equation}
where $q$ is the electric charge of the point particle. The equations governing the dynamics of the electromagnetic tensor $F_{\mu\nu}$ (that is, Maxwell equations) in presence of a gravitation field and the influence of electricity and magnetism on gravitation will be briefly discussed in Section III, part A. It should be mentioned that all the equations so far - including Maxwell's - can be derived from a variational principle. For details in the derivation, one could consult references \cite{weinberg,mis,web} and \cite{dirac}.

Our next task is to establish the weak field limit of the Einstein equations. We assume a $quasi$-Minkowskian metric, that is, a metric in the form
\begin{equation}
g_{\mu\nu} = \eta_{\mu\nu} + h_{\mu\nu}  \,,\label{e6}
\end{equation}
where $\eta_{\mu \nu}$ is the Minkowski (that is, flat space-time)  metric tensor, represented by the matrix
\begin{equation}
    \begin{split}
        \begin{bmatrix}
            \eta _{\mu \nu}
        \end{bmatrix}=\begin{bmatrix}
            1 & 0 & 0 & 0\\
            0 & -1 & 0 & 0\\
            0 & 0 & -1 & 0\\
            0 & 0 & 0 & -1
        \end{bmatrix} 
    \end{split}  \,\label{e7}
\end{equation}
and
\begin{equation}
|h_{\mu\nu}\ | << 1  \,.\label{e8}
\end{equation}
The quantities $h_{\mu\nu}$ will henceforth be called \textit{perturbations of the metric}. It can be shown \cite{weinberg} that in this linear limit the components of the Einstein tensor are given by
\begin{equation}
\begin{split}
G_{\mu\nu} = \frac{1}{2}  \left ( \Box h_{\mu\nu} - \eta_{\mu\nu}\Box h
 + \frac{\partial ^2 h}{\partial x^{\mu}\partial x^{\nu}} \right. \\
\left.  + \eta_{\mu\nu}\eta^{\rho\alpha}\eta^{\sigma\beta}
\frac{\partial ^2 h_{\alpha\beta}}{\partial x^{\rho} \partial x^{\sigma} } \right. \; \; \; \; \; \; \; \; \; \; \; \\
\left.  -\eta^{\rho\alpha}
\frac{\partial ^2 h_{\alpha\mu}}{\partial x^{\nu} \partial x^{\rho}} 
-\eta^{\rho\alpha}
\frac{\partial ^2 h_{\alpha\nu}}{\partial x^{\mu} \partial x^{\rho}} \right )
\,,\label{e9}
\end{split}
\end{equation}
where $h=\eta^{\mu\nu}h_{\mu\nu} = {h^\mu}_{\mu}$ is the trace of the tensor $h_{\mu\nu}$ and $\Box$ is the D'Alembert wave operator. Substituting equation (9) in the left hand side of equation (1) gives us the desired weak field equations.

It is well known that in general we are, to a certain extent, free to choose conveniently the properties of the coordinate system to be used in a specific physical problem. Particularly in dealing with gravitational waves in vacuum, usually the preferred one is the \textit{transverse-traceless} (or $TT$) system of coordinates \cite{mis}, in which the metric perturbations obeys both the transversalityt condition
\begin{equation}
\frac{\partial {h^\mu}_{\nu}}{\partial x^{\mu}} = 0  \, \label{e10}
\end{equation}
and the traceless condition
\begin{equation}
h = 0  \, .\label{e11}
\end{equation}
In fact, the $TT$ coordinates are a subcategory of harmonic coordinates, for which
\begin{equation}
\frac{\partial {h^\mu}_{\nu}}{\partial x^{\mu}} - \frac{1}{2} \frac{\partial h}{\partial x^{\nu}}= 0  \,. \label{e12}
\end{equation}
With conditions (10) and (11) enforced in equation (9) applied to empty space, equation (1) assumes the very simple and suggestive form
\begin{equation}
\Box h_{\mu\nu} = 0 \,.\label{e13}
\end{equation}
Field equation (13) gives us the impression that generally all ten independent components of the metric tensor can exhibit a wave-like behavior, which by no means corresponds to a physical reality. The solution of equation (13) for a plane gravitational wave propagating in a specific direction furnishes the well known $plus$ and $cross$ independent tensor polarization states for these waves, correctly showing that gravitational waves in vacuum just have two degrees of freedom associated solely to the space-space transverse components of the metric, the other components being zero \cite{mis}. Furthermore, we stress that the aforementioned $TT$ system of coordinates can be chosen only in vacuum, so the dynamics of the metric in a material medium - particularly in a plasma - can not be taken into account in this oversimplified formulation. Fortunately, all the difficulties pointed out above can be removed in a gauge-free version of the linear theory of general relativity, as we shall see below.

\subsection{Gauge freedom, gauge invariance, and Einstein field equations}

In electromagnetism we write field equations for invariant gauge quantities and express the electromagnetic force in terms of these same quantities: the invariant quantities are the vector fields $\mathbf{E}$ and $\mathbf{B}$, which satisfy Maxwell equations, and the force law is derived from the Lorentz force \cite{jack}. It is desirable, as far as possible, that the same can be done in the linear theory of gravity. Furthermore, it is worth mentioning that a theory following the same spirit for non-relativistic quantum plasmas has already been obtained \cite{haas}. In this theory, the gauge invariant Wigner function (GIWF) is taken as the basis of a fluid-like system describing the plasma.  So, our main objectives in this subsection are:

(i) to express Einstein field equations in the linear regime in terms of gauge invariant quantities and

(ii) to show that the only radiative gravitational modes, in any gauge and in presence of sources (inclusive), correspond precisely to the pure spatial part of $h_{\mu \nu}$ satisfying conditions of null trace and transversality.

With this, it is removed from the theory any kind of difficulty and obscurity involving the non-distinction between objective gravitational waves and mere artifices resulting from a given choice of coordinate system (which, in general relativity, is the same as a choice of gauge). We begin the discussion by establishing expressions for the components of the metric tensor which will allow us to find gauge invariant quantities in the linear theory as directly as possible. Except for conventions, for some notation and for brevity in our approach, our rationale in this subsection follows closely that of references \cite{flan} and \cite{inan}. Furthermore, a more general alternative approach - although more abstract - is found in reference \cite{garg1}.

The scalar part of the perturbation tensor – that is, the time-time component of $h_{\mu \nu}$, invariant under spatial rotations – can be identified with the Newtonian gravitational potential $\phi$ according to
\begin{equation}
h_{00} = \frac{2 \phi}{c^2} \,.\label{e14} 
\end{equation}
This identification goes back to the origins of general relativity, where the equations for the gravitational field were constructed based on the equivalence principle and forced to reproduce the results of the Newtonian theory in the limit of weak fields and bodies moving at low speeds \cite{weinberg}.

The vector part of $h_{\mu \nu}$ – that is, the part that transforms as a 3-vector under spatial rotations – is identified with the set of time-space type components of $h_{\mu \nu}$:
\begin{equation}
    \begin{split}
        \begin{bmatrix}
            h_{01} & h_{02} & h_{02}
        \end{bmatrix}=-\frac{1}{c}\begin{bmatrix}
            A_x & A_y & A_z
        \end{bmatrix}=-\frac{1}{c}\textbf{A}
    \end{split}  \,.\label{e15}
\end{equation}
We proceed a little further now by employing the Helmholtz decomposition for the vector \textbf{A}. The Helmholtz theorem states that the vector \textbf{A}, subject to the appropriate boundary condition at spatial infinity
\begin{equation}
\lim_{\mathbf{x} \rightarrow \infty}{\mathbf{A}} = 0  \,\label{e16}
\end{equation}
can be expressed as
\begin{equation}
\mathbf{A} = \nabla \psi + \mathbf{h}  \,,\label{e17}
\end{equation}
where $\psi$ is a specific scalar field and \textbf{h} is a vector field satisfying
\begin{equation}
\nabla \cdot \mathbf{h} = 0  \,.\label{e18}
\end{equation}
Equations (17) and (18) allows expressing equation (15) in terms of components as
\begin{equation}
h_{0i} = -\frac{1}{c} \left ( \frac{\partial \psi}{\partial x^i} + 
\textsf{h}_i \right )  \,.\label{e19}
\end{equation}
with the condition
\begin{equation}
\frac{\partial \textsf{h}^i}{\partial x^i} = 0  \,.\label{e20}
\end{equation}
The quantities $\textsf{h}_i$ are the components of the 3-vector $\mathbf{h}$, and so here the operations of raising and lowering indices must be performed by contraction with the Krönecker delta, as in $\textsf{h}^i = \delta^{ij} \textsf{h}_j$ (for 3-vectors, the lowering and raising indices operations must be performed with $\delta_{ij}$ and $\delta^{ij}$, respectively).

A strategy similar to the one outlined above can be employed in structuring the tensor part of $h_{\mu \nu}$ - that is, the part that transforms according to a 3-tensor in spatial rotations. It can be shown that this tensor part of the perturbation admits a Helmholtz decomposition of the form
\begin{equation}
\begin{split}
h_{ij} = & h^{TT}_{ij} + \frac{1}{3} \delta_{ij}H + \frac{1}{2} \left ( \frac{\partial \mathcal{E}_i}{\partial {x^j}} 
+ \frac{\partial \mathcal{E}_j}{\partial {x^i}}\right ) \\
& + \left ( \frac{\partial^2}{\partial x^i \partial x^j}
- \frac{1}{3}\delta_{ij}\nabla ^2
\right ) \lambda
\,,\label{e21}
\end{split}
\end{equation}
where $h^{TT}_{ij}$ is the \textit{transverse traceless} part of $h_{ij}$, satisfying the conditions
\begin{equation}
\frac{\partial h^i \, ^{TT}_j}{\partial x^i} = 0  \, \label{e22}
\end{equation}
\begin{equation}
\delta ^{ij} h^{TT}_{ij} = 0  \,.\label{e23}
\end{equation}
Here, $\lambda$ and $\mathcal{E}_i$ are, respectively, the scalar and vector potential for $h_{ij}$, the last one satisfying
\begin{equation}
\frac{\partial \mathcal{E}^i}{\partial x^i} = 0  \,.\label{e24}
\end{equation}
By contracting equation (21) with $\delta^{ij}$ and employing equations (23) and (24), we see that $H = \delta^{ij} h_{ij}$, that is, the (3-dimensional) trace of $h_{ij}$.

A second order symmetric 4-tensor has ten independent components. But calculating the total number of independent functions given by equations (14), (19) and (21) does not give this value. In fact, in the proposed decomposition we have four scalar functions ($\phi$, $\psi$, $H$ and $\lambda$), three vector components $\textsf{h}_i$, three vector components $\mathcal{E}_i$ and the six components of the symmetric 3-tensor $h^{TT}_{ij}$, giving sixteen functions. On the other hand, there are six constraints - one given by equation (20), three given by equations (22), one given by equation (23), plus one given by equation (24) – which gives us 16 – 6 = 10 independent functions, as expected. Note that equations (22) and (23) are closely related to the traceless transverse coordinate condition usually employed in the discussion of the propagation and properties of gravitational waves in free space. However, there is no gauge fixing here: all we have done is to establish an appropriate decomposition for the perturbation tensor $h_{\mu \nu}$, with the correct number of independent components (ten), without any kind of restriction on the physical system in question or on the adopted coordinate system (except, of course, that we are working in the linear regime of the field equations). It is also worth mentioning that it can be proved that the decomposition presented here is unique when the quantities $\psi$, $\lambda$ and $\mathcal{E}_i$ are subjected to certain fairly simple and reasonable physical universal boundary conditions. However,
we will not deal with the details of this matter here \cite{flan}.

Let us now look for invariant gauge quantities in the linear theory, in a sense the gravitational analogues of the $\mathbf{E}$ and $\mathbf{B}$ fields of electromagnetism. As is well known, in the linear regime the perturbations of the metric transforms, in a coordinate system change (or gauge transformation), as \cite{weinberg}
\begin{equation}
{h'}_{\mu \nu} = h_{\mu \nu}
-\frac{\partial \chi_\nu}{\partial x^\mu}
-\frac{\partial \chi_\mu}{\partial x^\nu} \,. \label{e25}
\end{equation}
Under spatial rotations, the transformation 4-vector $\chi_\mu$ can also be conveniently decomposed into a scalar and a vector part according to
\begin{equation}
\chi_\mu = \left ( \chi_0, \chi_i \right ) =
\left ( \frac{\eta}{c}, \frac{\partial a}{\partial x^i} + b_i \right )\,, \label{e26}
\end{equation}
with the condition
\begin{equation}
\frac{\partial b^i}{\partial x^i} = 0 \,. \label{e27}
\end{equation}
Note that in equations (26) and (27) above we employ Helmholtz's theorem again, now for the 3-vector $\chi_i$. Like $\chi_\mu$, the transformation functions $\eta$, $a$ and $b^i$ (subject to constraint (27)) are completely arbitrary \textit{except for keeping the field weak}.

For the scalar component of the perturbation tensor, the transformation (25) gives
\begin{equation}
\phi' = \phi - \frac{\partial\eta}{\partial t} \,. \label{e28}
\end{equation}
as can be verified by substituting equation (14) in the transformation (25) and employing equation (26). Substituting the components given in equation (19) into equation (25), we find after some algebra
\begin{equation}
\frac{\partial \psi'}{\partial x^i} + \textsf{h}'_i
= \frac{\partial}{\partial x^i} \left (\psi + \eta + \frac{\partial a}{\partial t} \right ) + \textsf{h}_i
+ \frac{\partial b_i}{\partial t}
 \,. \label{e29}
\end{equation}
Operating with $\delta^{ij} \frac{\partial}{\partial x^j}$ on equation (29) and taking into account the constraints (20) and (27), we find
\begin{equation}
\psi' = \psi + \eta + \frac{\partial a}{\partial t}
 \,, \label{e30}
\end{equation}
which, when inserted back into equation (29), gives
\begin{equation}
\textsf{h}'_i = \textsf{h}_i + \frac{\partial b_i}{\partial t} 
 \,. \label{e31}
\end{equation}
Finally, performing the transformation (25) of the pure spatial components of $h_{\mu \nu}$, given by equation (21), we now find
\begin{equation}
\begin{split}
h'^{TT}_{ij} + \frac{1}{3} \delta_{ij}H' + \frac{1}{2} \left ( \frac{\partial \mathcal{E}'_i}{\partial {x^j}} 
+ \frac{\partial \mathcal{E}'_j}{\partial {x^i}}\right ) \\
+ \left ( \frac{\partial^2}{\partial x^i \partial x^j}
- \frac{1}{3}\delta_{ij}\nabla ^2
\right ) \lambda' \\
= h^{TT}_{ij} + \frac{1}{3} \delta_{ij}H + \frac{1}{2} \left ( \frac{\partial \mathcal{E}_i}{\partial {x^j}} 
+ \frac{\partial \mathcal{E}_j}{\partial {x^i}}\right ) \\
+ \left ( \frac{\partial^2}{\partial x^i \partial x^j}
- \frac{1}{3}\delta_{ij}\nabla ^2
\right ) \lambda \\
- 2 \frac{\partial^2 a}{\partial x^i \partial x^j} - \frac{\partial b_i}{\partial x^j} - \frac{\partial b_j}{\partial x^i}
\,.\label{e32}
\end{split}
\end{equation}
Contracting equation (32) with $\delta^{ij}$ and taking into account the constraints (22), (23), (24) and (27), we find
\begin{equation}
H' = H - 2 \nabla^2a
 \,. \label{e33}
\end{equation}
Returning with the result (33) into equation (32) and rearranging, we obtain
\begin{equation}
\begin{split}
h'^{TT}_{ij} + & \frac{1}{2} \left ( \frac{\partial \mathcal{E}'_i}{\partial {x^j}} 
+ \frac{\partial \mathcal{E}'_j}{\partial {x^i}}\right ) + \left ( \frac{\partial^2}{\partial x^i \partial x^j}
- \frac{1}{3}\delta_{ij}\nabla ^2
\right ) \lambda' \\
= \; & h^{TT}_{ij} + \frac{1}{2} \left [ \frac{\partial \left (\mathcal{E}_i - 2b_i \right )}{\partial {x^j}} 
+ \frac{\partial \left (\mathcal{E}_j - 2b_j \right )}{\partial {x^i}}\right ] \\
& + \left ( \frac{\partial^2}{\partial x^i \partial x^j}
- \frac{1}{3}\delta_{ij}\nabla ^2
\right ) \left ( \lambda - 2a\right ) \\
\,\label{e34}
\end{split}
\end{equation}
By comparing the terms to the left and right sides of equation (34), we conclude that
\begin{equation}
h'^{TT}_{ij} = h^{TT}_{ij}\, \label{e35}
\end{equation}
\begin{equation}
\mathcal{E}'_i = \mathcal{E}_i -2b_i\, \label{e36}
\end{equation}
\begin{equation}
\lambda' = \lambda - 2a\,, \label{e37}
\end{equation}
which completes the picture of the behavior of the functions $\phi$, $\psi$, $H$, $\lambda$, $\textsf{h}_i$, $\mathcal{E}_i$, and $h^{TT}_{ij}$ under the gauge transformations of the linear gravitational theory. Note that, so far, the only invariant quantities are the tensor components $h^{TT}_{ij}$, just those related to gravitational radiation in the traditional approach using the transverse traceless gauge. From the set of transformations (28), (30), (31), (33), (35), (36) and (37) we can find two more scalar invariants and a vector invariant. The first scalar invariant, $\Phi$, is obtained by adding equation (28) with the first time derivative of equation (30) and with one half of the second time derivative of equation (37). That gives us
\begin{equation}
\Phi = \phi' + \frac{\partial \psi'}{\partial t}
+ \frac{1}{2} \frac{\partial^2 \lambda'}{\partial t^2} = \phi + \frac{\partial \psi}{\partial t} + \frac{1}{2} \frac{\partial^2
\lambda}{\partial t^2}
\,. \label{e38}
\end{equation}
The second scalar invariant, $\Theta$, is obtained by taking the Laplacian of equation (33) and subtracting equation (37), giving
\begin{equation}
\Theta = \frac{1}{3} \left (H' - \nabla^2 \lambda' \right ) = \frac{1}{3} \left (H - \nabla^2 \lambda \right )
\,. \label{e39}
\end{equation}
The factor $\frac{1}{3}$ in equation (39) has been included for convenience. Finally, a vector invariant $\Xi_i$ is obtained by adding equation (31) with one half of the first time derivative of equation (36), which results in
\begin{equation}
\Xi_i = \textsf{h}'_i + \frac{1}{2} \frac{\partial \mathcal{E}'_i}{\partial t} = \textsf{h}_i + \frac{1}{2} \frac{\partial \mathcal{E}_i}{\partial t}
\,. \label{e40}
\end{equation}
Note that, due to divergenceless conditions (20) and (24), the vector $\Xi_i$ satisfies
\begin{equation}
\frac{\partial \Xi^i}{\partial x^i} = 0
\,. \label{e41}
\end{equation}
Our next goal is to express the metric tensor in terms of the eleven invariant quantities $\Phi$, $\Theta$, $\Xi_i$ and $h^{TT}_{ij}$ and to establish the equations of motion for these quantities. In fact, considering the five constraints given by equations (22), (23) and (41), of these quantities we have just 11 – 5 = 6 independent functions, meaning that of the ten components of the tensor $h_{\mu \nu}$, in general only six are physically significant.

By isolating the functions $\phi$, $H$ and $\textsf{h}_i$ respectively in equations (38), (39) and (40), and adequately substituting the results in equations (14), (19) and (21), we can show that the scalar, vector and tensor components of the perturbation tensor $h_{\mu \nu}$ can be expressed in terms of gauge invariants in the form:
\begin{equation}
h_{00} = \frac{2}{c^2} \left (\Phi - \frac{\partial \psi}{\partial t} - \frac{1}{2} \frac{\partial^2 \lambda}{\partial t^2} \right )
\, \label{e42}
\end{equation}
\begin{equation}
h_{0i} = -\frac{1}{c} \left (\Xi_i + \frac{\partial \psi}{\partial x^i} - \frac{1}{2} \frac{\partial \mathcal{E}_i}{\partial t} \right )
\, \label{e43}
\end{equation}
\begin{equation}
h_{ij} = h^{TT}_{ij} + \delta_{ij}\Theta
+ \frac{1}{2} \left ( \frac{\partial \mathcal{E}_i}{\partial {x^j}} 
+ \frac{\partial \mathcal{E}_j}{\partial {x^i}}\right ) + \frac{\partial^2 \lambda}{\partial x^i \partial x^j}
\, \label{e44}
\end{equation}
To clarify: the metric perturbation $h_{\mu \nu}$ depends on the eleven invariants $\Phi$, $\Theta$, $\Xi_i$ and $h^{TT}_{ij}$ subjected to the five conditions (22), (23) and (41) – giving six physical degrees of freedom, as we saw – and on the five gauge-dependent quantities $\psi$, $\lambda$ and $\mathcal{E}_i$ subjected to the condition (24), totaling 5 – 1 = 4 gauge degrees of freedom. The total number of degrees of freedom is 6 + 4 = 10, as expected for a symmetric second-order 4-tensor.

By equations (42)–(44) we see that it is not possible to express the metric tensor only in terms of invariant gauge quantities, and this conclusion reflects the equivalence principle: at any point in space-time it is always possible to find a locally inertial frame of reference, which can be mathematically achieved with appropriate choices of the functions $\psi$, $\lambda$ and $\mathcal{E}_i$.

It can be shown from equation (9) that, despite the gauge dependence of the metric, the components of the Einstein tensor $G_{\mu \nu}$ \textit{are} expressed only in terms of the gauge invariant quantities, as given below:
\begin{equation}
G_{00} = -\nabla^2 \Theta
\, \label{e45}
\end{equation}
\begin{equation}
G_{0i} = -\frac{1}{c} \left ( \frac{\partial^2 \Theta}{\partial t \partial x^i} - \frac{1}{2}\nabla^2 \Xi_i \right )
\, \label{e46}
\end{equation}
\begin{equation}
\begin{split}
& G_{ij} = \frac{1}{2} \Box h^{TT}_{ij}
- \delta_{ij} \frac{1}{c^2} \frac{\partial^2 \Theta}{\partial t^2} \, \\
& + \frac{1}{2c^2} \left ( \frac{\partial}{\partial x^j} \frac{\partial \Xi_i}{\partial t}
+\frac{\partial}{\partial x^i} \frac{\partial \Xi_j}{\partial t}
\right ) \\
& + \frac{1}{2} \left (\frac{\partial^2}{\partial x^i \partial x^j} - \delta_{ij} \nabla^2 \right ) \left (  \frac{2\Phi}{c^2} - \Theta \right )
\, \label{e47}
\end{split}
\end{equation}
By checking equation (47) we see that the wave operator $\Box$ appears just in the first term at the right side, acting upon the transverse traceless part of the metric perturbation. As we will see soon, this implies that these components are the only obeying a wave equation and so are radiative, corresponding to the wave modes of the gravitational field.

The Einstein field equations (1) can now be split in a series of simpler equations, one for each of the gauge invariant quantities $\Phi$, $\Theta$, $\Xi_i$ and $h^{TT}_{ij}$. The equation for $\Theta$ is readily obtained simply by taking the time-time component of equation (1), that is, by writing
\begin{equation}
G_{00} = -\frac{8\pi G}{c^4} T_{00} \,. \label{e48}
\end{equation}
From equation (45) substituted in equation (48) above, we obtain
\begin{equation}
\nabla^2 \Theta = \frac{8\pi G}{c^4} T_{00}
\, . \label{e49}
\end{equation}
Next, to establish an equation for the gauge invariant vector $\Xi_i$, we take the time-space type components of equation (1), getting
\begin{equation}
G_{0i} = -\frac{8\pi G}{c^4} T_{0i} \,. \label{e50}
\end{equation}
For the left hand side of equation (50) we substitute equation (46), and for the right side we employ a Helmholtz decomposition of the form  
\begin{equation}
T_{0i} = c \left ( \frac{\partial S}{\partial x^i} + S_i \right )  \,,\label{e51}
\end{equation}
with the condition
\begin{equation}
\frac{\partial S^i}{\partial x^i} = 0  \,.\label{e52}
\end{equation}
So, the equation (50) can now be written as
\begin{equation}
\frac{\partial^2 \Theta}{\partial t \partial x^i} - \frac{1}{2}\nabla^2 \Xi_i = \frac{8\pi G}{c^2}  \left ( \frac{\partial S}{\partial x^i} + S_i \right ) \,.\label{e53}
\end{equation}
It can be shown that, by taking the 3-divergence of equation (53), we can recover the result (49). On the other hand, we can readily put equation (53) in the explicit vector form
\begin{equation}
\nabla \frac{\partial \Theta}{\partial t} - \frac{1}{2} \nabla^2 \mathbf{\Xi}
= \frac{8\pi G}{c^2} \left ( \nabla S + \textbf{S} \right )
\,.\label{e54}
\end{equation}
Then, by taking the curl of equation (54) and, remembering that the curl of any gradient is zero, we are led to
\begin{equation}
\nabla \times \left ( \frac{1}{2} \nabla^2 \mathbf{\Xi}
+ \frac{8\pi G}{c^2} \textbf{S} \right ) = 0
\,.\label{e55}
\end{equation}%
Equation (55) implies that
\begin{equation}
\frac{1}{2} \nabla^2 \mathbf{\Xi}
+ \frac{8\pi G}{c^2} \textbf{S} = \nabla f(\mathbf{x},t)
\,, \label{e56}
\end{equation}
where $f(\mathbf{x},t)$ is some scalar function of coordinates and time. Since the 3-vectors $\mathbf{\Xi}$ and \textbf{S} are both divergenceless and are assumed to vanish as $\mathbf{x} \rightarrow \infty$, equation (56) implies that
\begin{equation}
\nabla^2 \mathbf{\Xi}
= - \frac{16\pi G}{c^2} \textbf{S}
\,,\label{e57}
\end{equation}
or, in component form,
\begin{equation}
\nabla^2 \Xi_i
= - \frac{16\pi G}{c^2} S_i
\,.\label{e58}
\end{equation}
Equation (58) is the differential equation for the gauge invariant vector $\Xi_i$. It is worth to mention that the equations of motion for $\Theta$ and for $\Xi_i$ (equations (49) and (58)) are both inhomogeneous Poisson-like equations (not wave-like), and so have only (non radiative) solutions decaying with $r^{-2}$, $r$ being a radial coordinate. Thus, neither $\Theta$ nor $\Xi_i$ can be associated with gravitational waves. Furthermore, notice that the source for $\Theta$ is $T_{00}$ and for $\Xi_i$ is just the \textit{rotational} part of $T_{0i}$, given by $S_i$.

To obtain equations of motion for the $\Phi$ and  $h^{TT}_{ij}$, we must take into account the space-space components of equation (1). Substituting equation (47) in the right side of equation (1) we obtain
\begin{equation}
\begin{split}
\frac{1}{2} \Box h^{TT}_{ij}
- \delta_{ij} \frac{1}{c^2} \frac{\partial^2 \Theta}{\partial t^2} + \frac{1}{2c^2} \left ( \frac{\partial}{\partial x^j} \frac{\partial \Xi_i}{\partial t}
+\frac{\partial}{\partial x^i} \frac{\partial \Xi_j}{\partial t}
\right ) \\
+ \frac{1}{2} \left (\frac{\partial^2}{\partial x^i \partial x^j} - \delta_{ij} \nabla^2 \right ) \left (  \frac{2\Phi}{c^2} - \Theta \right ) = - \frac{8\pi G}{c^4} T_{ij}
\, \label{e59}
\end{split}
\end{equation}
To find the differential equation to $\Phi$, we contract equation (59) with $\delta^{ij}$ and use the conditions given by equations (23) and (41). We obtain, after some algebraic manipulations,
\begin{equation}
\frac{1}{c^2} \frac{\partial^2 \Theta}{\partial t^2} + \frac{2}{3c^2}  \nabla^2 \Phi - \frac{1}{3} \nabla^2\Theta =  \frac{8\pi G}{c^4} P
\,, \label{e60}
\end{equation}
where $P$ is given by
\begin{equation}
\begin{split}
P = \frac{1}{3} \delta^{ij}T_{ij}
\, \label{e61}
\end{split}
\end{equation}
Substituting equation (49) for $\nabla^2 \Theta$ in the third term at left side in equation (60) ant taking the Laplacian of the resulting equation, we get
\begin{equation}
\nabla^2 \frac{\partial^2 \Theta}{\partial t^2} + \frac{2}{3} \nabla^2 \nabla^2 \Phi - \frac{8\pi G}{3 c^2} \nabla^2 T_{00} = \frac{8\pi G}{c^2} \nabla^2 P
\,. \label{e62}
\end{equation}
The first term at left side in equation (62) may be tamed by taking the divergence of equation (54) and employing the divergenceless conditions (41) and (52). The result is
\begin{equation}
\nabla^2 \frac{\partial \Theta}{\partial t} = \frac{8\pi G}{c^2} \nabla^2 S \,.\label{e63}
\end{equation}
Now, taking the time derivative of equation (63), it gives
\begin{equation}
\nabla^2 \frac{\partial^2 \Theta}{\partial t^2} = \nabla^2 \left ( \frac{8\pi G}{c^2} \frac{\partial S}{\partial t} \right ) \,.\label{e64}
\end{equation}
Returning with the result (64) in equation (62) and rearranging we obtain
\begin{equation}
\nabla^2 \left ( \frac{4\pi G}{c^2} \frac{\partial S}{\partial t} + \frac{1}{3} \nabla^2 \Phi - \frac{4\pi G}{3 c^2} T_{00} - \frac{4\pi G}{c^2} P \right ) = 0 \,. \label{e65}
\end{equation}
Again, since the functions $S$, $\Phi$, $T_{00}$ and $P$ are all assumed to vanish as $\mathbf{x} \rightarrow \infty$, equation (65) implies that
\begin{equation}
\frac{4\pi G}{c^2} \frac{\partial S}{\partial t} + \frac{1}{3} \nabla^2 \Phi - \frac{4\pi G}{3 c^2} T_{00} - \frac{4\pi G}{c^2} P = 0 \,, \label{e66}
\end{equation}
that is,
\begin{equation}
\nabla^2 \Phi = \frac{4\pi G}{c^2} \left [ T_{00} + 3 \left (P - \frac{\partial S}{\partial t} \right ) \right ] \,. \label{e67}
\end{equation}
which is the differential equation for $\Phi$ – again, a Poisson equation.

Finally, the differential equation for $h^{TT}_{ij}$ will be found, again by properly manipulating the equation (59). However, at this point of the development, it is useful first to exploit the conservation law
\begin{equation}
\frac{\partial {T^\mu}_{\nu}}{\partial x^\mu} = 0 \;, \label{e68}
\end{equation}
which is valid in this form to the first order. For $\nu = 0$ we obtain from equation (68) an equation for energy conservation in the form
\begin{equation}
\frac{1}{c^2} \frac{\partial T_{00}}{\partial t} = \nabla^2 \mathbf{S} \;, \label{e69}
\end{equation}
where use was made of equations (51) and (52). To exploit the case with $\nu = i$, it is convenient to express the components of the stress tensor $T_{ij}$ in the Helmholtz decomposed form
\begin{equation}
\begin{split}
T_{ij} = & \; \sigma_{ij} + \delta_{ij}P
+ \left ( \frac{\partial \sigma_i}{\partial x^j} + \frac{\partial \sigma_j}{\partial x^i} \right ) \\
& + \left ( \frac{\partial^2}{\partial x^i \partial x^j} - \frac{1}{3} \delta_{ij} \nabla^2 \right ) \sigma
\;, \label{e70}
\end{split}
\end{equation}
with
\begin{equation}
\frac{\partial {\sigma^i}_j}{\partial x^i} = 0 \; \label{e71}
\end{equation}
\begin{equation}
\delta^{ij} \sigma_{ij} = 0
\; \label{e72}
\end{equation}
\begin{equation}
\frac{\partial \sigma^i}{\partial x^i} = 0 \;. \label{e73}
\end{equation}
By virtue of equations (71) and (72), the $\sigma_{ij}$ part of $T_{ij}$ is called the \textit{transverse traceless} part of the stress tensor. Now, setting $\nu = i$ in equation (68) and taking into account equations (51) and (70), we obtain a conservation equation that can be readily written in the explicit vector form
\begin{equation}
\nabla \frac{\partial S}{\partial t}
+ \frac{\partial \mathbf{S}}{\partial t}
= \nabla P + \nabla^2 \mathbf{\Sigma} + \frac{2}{3} \nabla^2 \nabla \sigma
\;, \label{e74}
\end{equation}
with
$
\mathbf{\Sigma} =
    \begin{bmatrix}
        \sigma_1 & \sigma_2 & \sigma_3
        \end{bmatrix}
$.
Taking the divergence of equation (74) we obtain
\begin{equation}
\nabla^2 \left [ \frac{3}{2} \left ( \frac{\partial S}{\partial t} - P \right ) - \nabla^2 \sigma \right ] = 0 \;, \label{e75}
\end{equation}
where use was made of conditions (52) and (71). On the other hand, taking the curl of equation (74) and using the fact that the curl of a gradient is always zero, we get
\begin{equation}
\nabla \times \left[ \frac{\partial \mathbf{S}}{\partial t} - \nabla^2 \mathbf{\Sigma} \right ] = 0 \;. \label{e76}
\end{equation}
Imposing the universal boundary conditions $S = P = \sigma = S_i = \sigma_i = 0$ as $\mathbf{x} \rightarrow \infty$, equations (75) and (76) gives, respectively,
\begin{equation}
\frac{3}{2} \left ( \frac{\partial S}{\partial t} - P \right ) = \nabla^2 \sigma \; \label{e77}
\end{equation}
and
\begin{equation}
\frac{\partial S_i}{\partial t} = \nabla^2 \sigma_i \;. \label{e78}
\end{equation}
Yet, from equation (77) we can write
\begin{equation}
\frac{\partial S}{\partial t} = \frac{2}{3} \nabla^2 \sigma + P \;, \label{e79}
\end{equation}
which will be used latter. Now, by taking the Laplacian of equation (59) we find
\begin{equation}
\begin{split}
\frac{1}{2} \nabla^2 \Box h^{TT}_{ij}
- \delta_{ij} \frac{1}{c^2} \nabla^2 \frac{\partial^2 \Theta}{\partial t^2} \; \; \; \; \; \; \; &\\
+ \frac{1}{2c^2} \left ( \frac{\partial}{\partial x^j} \frac{\partial \nabla^2 \Xi_i}{\partial t}
+\frac{\partial}{\partial x^i} \frac{\partial \nabla^2 \Xi_j}{\partial t} \right ) & \\
+ \frac{1}{2} \left (\frac{\partial^2}{\partial x^i \partial x^j} - \delta_{ij} \nabla^2 \right ) \nabla^2 \left (  \frac{2\Phi}{c^2} - \Theta \right ) + \frac{8\pi G}{c^4} & \nabla^2 T_{ij} = 0
\,. \label{e80}
\end{split}
\end{equation}
Substituting equations (49), (58), (64) and (67), respectively, for $\nabla^2 \Theta$, $\nabla^2 \Xi_i$, $\nabla^2 \frac{\partial^2 \Theta}{\partial t^2}$ and $\nabla^2 \Phi$ in equation (80) we find, after rearranging, 
\begin{equation}
\begin{split}
- \frac{c^4}{16\pi G} \nabla^2 \Box h^{TT}_{ij}
+ \delta_{ij} \nabla^2 \frac{\partial S}{\partial t}
+ \left ( \frac{\partial}{\partial x^j} \frac{\partial S_i}{\partial t}
+ \frac{\partial}{\partial x^i} \frac{\partial S_j}{\partial t}
\right ) \\
+ \left ( \frac{\partial^2}{\partial x^i \partial x^j} - \delta_{ij} \nabla^2
\right ) \frac{3}{2} \left ( \frac{\partial S}{\partial t} - P \right ) - \nabla^2 T{ij}
= 0
\,. \label{e81}
\end{split}
\end{equation}
At this point, substituting equations (77), (78) and (79), respectively, for $\frac{3}{2} \left (\frac{\partial S}{\partial t} - P \right )$, $\frac{\partial S_i}{\partial t}$ and $\frac{\partial S}{\partial t}$ in equation (81) and grouping similar terms, we are led to
\begin{equation}
\begin{split}
- \frac{c^4}{16\pi G} \nabla^2 \Box & h^{TT}_{ij}
+ \delta_{ij} \nabla^2 P 
+ \left (\frac{\partial^2}{\partial x^i \partial x^j} - \frac{1}{3} \delta_{ij} \nabla^2 \right ) \nabla^2 \sigma \\
& + \left ( \frac{\partial \nabla^2 \sigma_i}{\partial x^j} + \frac{\partial \nabla^2 \sigma_j}{\partial x^i}
\right ) - \nabla^2 T_{ij} = 0
\,. \label{e82}
\end{split}
\end{equation}
Finally, by the Helmholtz decomposition for $T_{ij}$, equation (70), we recognize that the second, third, fourth and fifth terms at left side in equation (82) when added gives exactly $-\sigma_{ij}$, that is
\begin{equation}
\frac{c^4}{16\pi G} \nabla^2 \Box h^{TT}_{ij} + \nabla^2 \sigma_{ij} = 0
\, \label{e83}
\end{equation}
or
\begin{equation}
\nabla^2 \left (  \Box h^{TT}_{ij}
+ \frac{16\pi G}{c^4} \sigma_{ij} \right ) = 0
\,. \label{e84}
\end{equation}
Assuming, as usual, that both $h^{TT}_{ij}$ and $\sigma_{ij}$ goes to zero as $\mathbf{x} \rightarrow \infty$, equation (84) implies that
\begin{equation}
\Box h^{TT}_{ij} = - \frac{16\pi G}{c^4} \sigma_{ij} \,. \label{e85}
\end{equation}
completing our set of equations for the gauge invariants $\Phi$, $\Theta$, $\Xi_i$ and $h^{TT}_{ij}$. Summarizing, in the weak field approximation of general relativity, the Einstein equations are split up in a set of linear inhomogeneous differential equations given by
\begin{equation}
\nabla^2 \Theta = \frac{8\pi G}{c^4} T_{00} \, \label{86}
\end{equation}
\begin{equation}
\nabla^2 \Phi = \frac{4\pi G}{c^2} \left [ T_{00} + 3 \left (P - \frac{\partial S}{\partial t} \right ) \right ] \, \label{e87}
\end{equation}
\begin{equation}
\nabla^2 \Xi_i
= - \frac{16\pi G}{c^2} S_i
\,\label{e88}
\end{equation}
\begin{equation}
\Box h^{TT}_{ij} = - \frac{16\pi G}{c^4} \sigma_{ij} \,, \label{e89}
\end{equation}
where $T_{00}$, $P$, $S$, $S_i$ and $\sigma_{ij}$ are related to the stress-energy tensor of the physical system under consideration, and represents the sources for the gravitational field. Solving equations (86)-(89) completely determine the six physical degrees of freedom of the metric. The metric perturbation itself, as discussed previously, are given by equations (42)-(44) and are clearly gauge dependent, as it has to be. The Christoffel symbols appearing in the equations of motion (3) and (5) are also gauge dependent, reflecting the equivalence principle. That is, the objects $\Gamma^\mu_{\nu \rho}$ can be locally made to vanish by an appropriate choice of coordinates. Chosen a gauge (or a system of coordinates) with $\psi = \lambda = \mathcal{E}_i = 0$ equations (42)-(44) assumes the very simple form
\begin{equation}
    h_{00} = \frac{2 \Phi}{c^2}
\, \label{e90}
\end{equation}
\begin{equation}
    h_{0i} = - \frac{1}{c} \Xi_i
\, \label{e91}
\end{equation}
\begin{equation}
    h_{ij} = h^{TT}_{ij} + \delta_{ij} \Theta
\,, \label{e92}
\end{equation}
and this will be our preferred gauge in explicit calculations involving the metric tensor. The Einstein field equations (86)-(89), of course, are not affected by this or any other choice of gauge, as they are written only in terms of gauge invariant quantities. Just to mention some interesting features of field equations (86)-(89) and the metric (90)-(92), note that in free space, all field equations are homogeneous and so the equations (86) and (87) are the same, leading to $\Theta = \frac{2 \Phi}{c^2}$. This identification is closely related to the Schwarzschild solution expressed in isotropic coordinates and correct to the first order, in which $h_{00} = \frac{2 \Phi}{c^2}$, $h_{ij} = \delta_{ij} \frac{2 \Phi}{c^2}$ and $h^{TT}_{ij} = h_{0i} = 0$, where $\Phi = -\frac{GM}{r}$ is the Newtonian gravitational potential of a point source of mass $M$. Furthermore, observe that equation (89) is the only wave equation of the set of field equations (86)-(89). This means that only the $h^{TT}_{ij}$ components of the metric perturbation behaves as waves and can be correctly identified as the radiative degrees of freedom of the gravitational field, as mentioned before. For the sake of completeness, in the next subsection this aspect will be briefly revised.

\subsection{Gravitational waves in free space}

It was pointed out in the subsection B of the present Section that gravitational radiation is associated to the $h^{TT}_{ij}$ part of the metric perturbation, which satisfies the traceless condition (22) and the divergenceless condition (23). In analogy with the wave solutions for electric and magnetic fields in electromagnetism, the general solution of the wave equation (89) is
\begin{equation}
    h^{TT}_{ij} (\mathbf{x},t) = -\frac{4 G}{c^4} \int \frac{\sigma_{ij} (\mathbf{x}',t')}{|\mathbf{x} - \mathbf{x}'|} d^3 x'
\,, \label{e93}
\end{equation}
where we use the definition of the retarded time, $t' = t - \frac{|\mathbf{x} - \mathbf{x}'|}{c}$ and the integration is performed over the source. Furthermore, in equation (93) it is understood that  $\sigma_{ij}$ are taken to the leading order. To the solutions given in equation (93), which describes the generation of the field by a gravitational source manifested in $\sigma_{ij} (\mathbf{x}', t')$, we can always add the solution of the homogeneous equation associated to equation (89)
\begin{equation}
    \Box h^{TT}_{ij} = 0
\,, \label{e94}
\end{equation}
satisfying to the conditions (22) and (23). Equation (94) admits as solutions plane waves of the form
\begin{equation}
    h^{TT}_{ij} = \epsilon_{ij} e^{i(\mathbf{k} \cdot \mathbf{x} - \omega t)} 
\,, \label{e95}
\end{equation}
where $\epsilon_{ij}$ are the components of the polarization 3-tensor, which is a symmetric tensor. By imposing the condition (22) to the plane wave solution (95) we find
\begin{equation}
    \epsilon_{ij}k^i = \epsilon_{1j}k^1 + \epsilon_{2j}k^2 + \epsilon_{3j}k^3 = 0 
\,. \label{e96}
\end{equation}
Furthermore, the traceless condition (23) implies that
\begin{equation}
    \epsilon_{11} + \epsilon_{22} + \epsilon_{33} = 0 
\,. \label{e97}
\end{equation}
As in the discussion of plane electromagnetic waves, an understanding of the features of the solution given in equations (95)-(97) is facilitated when considering the propagation along a specific coordinate axis, say $z$ axis, in the direction of increasing $z$. In this case, $k^1 = k^2 = 0$ and equation (96) furnishes $\epsilon_{3j}k^3 = 0$, that is,
\begin{equation}
    \epsilon_{31} = \epsilon_{32} = \epsilon_{33} = 0 
\,, \label{e98}
\end{equation}
meaning that every component $\epsilon_{ij}$ related to the direction of propagation (the $z$ direction) has to be zero. This shows that, indeed, gravitational waves are transverse. With $\epsilon_{33} = 0$, from equation (97) we get
\begin{equation}
    \epsilon_{22} = - \epsilon_{11} 
\,. \label{e99}
\end{equation}
Finally, collecting all the results (95), (98) and (99) together in matrix form we obtain
\begin{equation}
    \begin{split}
        \begin{bmatrix}
            h^{TT}_{ij}
        \end{bmatrix}=\begin{bmatrix}
            \epsilon_{11} & \epsilon_{12} & 0\\
            \epsilon_{12} & - \epsilon_{11} & 0\\
            0 & 0 & 0
        \end{bmatrix} e^{i(kz - \omega t)} \\
        =
        \begin{bmatrix}
            h_{+} & 0 & 0\\
            0 & - h_{+} & 0\\
            0 & 0 & 0
        \end{bmatrix} +
        \begin{bmatrix}
            0 & h_{\times} & 0\\
            h_{\times} & 0 & 0\\
            0 & 0 & 0
        \end{bmatrix}
    \end{split}  \, \label{e100}
\end{equation}
The two polarization states of gravitational waves are represented by the two matrices at right side in equation (100): the “plus” polarization are associated to the function $h_{+}(z,t) = \epsilon_{11}e^{i(kz - \omega t)}$ and the “cross” polarization to $h_{\times}(z,t) = \epsilon_{12}e^{i(kz - \omega t)}$. So, there are two different gravitational wave modes (transverse and mutually independent) contained in $h_{ij}$, oscillating in the $xy$ plane. In analogy with the electromagnetic case, the values of the amplitudes $\epsilon_{11}$ and $\epsilon_{12}$ completely specify the field.

Concluding this subsection, it is worth to mention that the result given in equation (100) are independent of any gauge, being just firmly based in the Helmholtz decomposition in the way that the tensor components $h^{TT}_{ij}$ must be taken as real physical objects. Furthermore, as the traceless and divergenceless (or transversality) conditions (22) and (23) imposed to $h^{TT}_{ij}$ are not related to a gauge choice for waves propagating in free space (as in the traditional approach using the Lorentz gauge), these conditions surely apply to gravitational waves propagating in material media, especially in a plasma.

\subsection{The energy carried by the gravitational field}

One of the most profound aspects of the Lagrangian formulation of field theory is the way conservation laws arises. From an appropriate Lagrangian density $\mathcal{L}$, the equations of motion for all fields of a physical system can be readily derived by imposing that the action $S$ obeys the \textit{variational principle}
\begin{equation}
\delta S = \delta \int d^4 x \sqrt{-g} \mathcal{L} = 0
\,, \label{e101}
\end{equation}
where $g$ is the determinant of the metric tensor. According to Noether's theorem \cite{sche}, if the action $S$ of a system is invariant under certain continuous transformation of the space-time coordinates and physical fields, then there will be a conserved quantity associated with this transformation (the transformation itself is called a \textit{symmetry}). Consider then a system composed by the gravitational field plus other fields (e.g. the electromagnetic and matter fields). When the symmetry of the action of this system under space-time translations is properly analyzed and established, then we are led to a momentum-energy conservation law. One of the mentioned conservation laws that can be derived from this procedure is
\begin{equation}
\frac{\partial}{\partial x^\mu}
\left [ \sqrt{-g} \left ( {t^\mu}_{\nu} + {T^\mu}_{\nu} \right ) \right ] = 0
\,, \label{e102}
\end{equation}
where ${T^\mu}_{\nu}$ are the mixed components of the stress-energy tensor of the system (including electromagnetic and matter terms, but not gravity) and ${t^\mu}_{\nu}$ is a gravitational stress-energy \textit{pseudo-tensor} given by
\begin{equation}
{t^\mu}_{\nu} = \frac{c^4}{16\pi G} \frac{1}{\sqrt{-g}} \left [ \left ( \Gamma^{\mu}_{\alpha \beta} - \delta^{\mu}_\beta \Gamma^{\gamma}_{\alpha \gamma} \right ) \frac{\partial \left ( \sqrt{-g} g^{\alpha \beta} \right ) }{\partial x^{\nu}} - \delta^{\mu}_{\nu} \bar{R}
\right ]
\,, \label{e103}
\end{equation}
with
\begin{equation}
\bar{R} = g^{\mu \nu} \left ( \Gamma^{\sigma}_{\mu \nu} \Gamma^{\rho}_{\rho \sigma}
- \Gamma^{\sigma}_{\mu \rho} \Gamma^{\rho}_{\nu \sigma}
\right )
\,. \label{e104}
\end{equation}
The form given for ${t^\mu}_{\nu}$ in equation (103)  and the related conservation law, equation (102), were obtained following Dirac's approach \cite{dirac},  although several other explicit forms can be chosen (e.g. the Landau-Lifshitz pseudo-tensor and related conservation law). For reasons of scope and space, we will not concern ourselves here with the derivation of the equations (102) and (103), neither make an effort to clarify why the gravitational stress-energy tensor is, in fact, a pseudo-tensor and why it is non-unique. For in-depth discussions of the subject, the reader can consult references \cite{dirac}, \cite{leod} and \cite{chiang}. The point that really must be clear hereafter is: because of the invariance of the action $S$ under space-time translations, the conservation of the \textit{total} energy and momentum (gravitation plus other fields) is guaranteed by equation (102), with gravitational contributions to the total stress-energy tensor given by equation (103). Observe that the leading term in ${t^\mu}_{\nu}$ is of second order, which justifies the conservation law given in equation (68), correct to the first order. It is important to emphasize that the entire discussion that we will undergo later on about the energy exchanges between matter and the gravitational field could also be carried out using other choices of the pseudo-tensor, leading to physically equivalent results. We chose the expressions given by equations (102) and (103) because they follow naturally (or, at least, more naturally) from the variational principle, which, in turn, provides a unified way of investigating physical systems (although we recognize that, in part, these choices were made for reasons of personal taste considering the background of the authors in their studies of general relativity).

Taking the covariant divergence of Einstein equation (equation (1)) and using that $\nabla_{\mu} {T^\mu}_{\nu} = 0$, it can be shown from equation (102) that
\begin{equation}
\frac{\partial \left ( \sqrt{-g} {t^\mu}_{\nu} \right )}{\partial x^{\mu}}
= \frac{c^4}{16\pi G} \sqrt{-g}
\frac{\partial g_{\rho \sigma}}{\partial x^\nu} G^{\rho \sigma}
\,. \label{e105}
\end{equation}
Especially, with $\nu = 0$, equation (105) becomes the (nearest of a) gravitational analogous of the Poynting theorem of electromagnetism. Indeed, identifying
\begin{equation}
 \sqrt{-g} {t^0}_0 = \mathcal{U}_g
\, \label{e106}
\end{equation}
and
\begin{equation}
c \sqrt{-g} {t^i}_0 = \mathcal{J}^i_g
\, \label{e107}
\end{equation}
respectively as the gravitational energy density and the gravitational energy flux, equation (105) is rewritten as
\begin{equation}
\frac{\partial \mathcal{U}_g}{\partial t}
+ \frac{\partial \mathcal{J}^i_g}{\partial x^i} 
= \frac{c^4}{16\pi G} \sqrt{-g}
\frac{\partial g_{\mu \nu}}{\partial t} G^{\mu \nu}
\,. \label{e108}
\end{equation}
Furthermore, considering the weak field approximation, equation (108) acquires the form
\begin{equation}
\frac{\partial \mathcal{U}_g}{\partial t}
+ \frac{\partial \mathcal{J}^i_g}{\partial x^i} 
= \frac{c^4}{16\pi G} \frac{\partial h_{\mu \nu}}{\partial t}
G^{\mu \nu}
\,, \label{e109}
\end{equation}
with $G^{\mu \nu}$ being the contravariant components of the Einstein tensor, given by equation (9). For gravitational waves in free space (or for any gravitational field in free space), where $G^{\mu \nu} = 0$ from the Einstein field equations (1), equations (108) and (109) guarantees gravitational energy conservation. Just to clarify the picture, we mention that, in the weak field approximation, equations (106) and (107) for energy density and energy flux of a plane gravitational wave propagating in the $z$ direction in vacuum gives
\begin{equation}
 <\mathcal{U}_g > \; = \frac{c^2 \omega^2}{32\pi G}
 \left [ \left (\epsilon_{11} \right )^2 + \left (\epsilon_{12} \right )^2 \right]
\, \label{e110}
\end{equation}
and
\begin{equation}
 <\mathcal{J}_g > \; = \frac{c^3 \omega^2}{32\pi G}
 \left [ \left (\epsilon_{11} \right )^2 + \left (\epsilon_{12} \right )^2 \right]
\,, \label{e111}
\end{equation}
where the symbol $<>$ represents the mean value over a period and $\omega = c \, k$ is the angular frequency of the wave. It is nice to observe that the results given in equations (110) and (111) greatly resembles that for electromagnetic waves. On the other hand, for a medium other than empty space, the right side of equations (108) and (109) are associated to the coupling between gravity and the other constituents of the physical system under consideration, giving the rate of exchange of gravitational energy. So, equations (108) and (109) constitutes suitable tools to study these exchanges.

Our last task in this Section is to furnish some useful formulae to calculate the Christoffel symbols in terms of the gauge invariant quantities obtained in part B of this Section (they are needed, among other things, to calculate the densities $\mathcal{U}_g$ and $\mathcal{J}^i_g$). For this purpose, we employ our preferred gauge with $\psi = \lambda = \mathcal{E}_i = 0$ and, with this, from equation (4) we obtain the complete list of Christoffel symbols of the second kind, below:
\begin{equation}
\Gamma^0_{00} = \frac{1}{c^3} \frac{\partial \Phi}{\partial t}
\label{e112}
\end{equation}
\begin{equation}
\Gamma^0_{0i} = \Gamma^0_{i0} = \frac{1}{c^2} \frac{\partial \Phi}{\partial x^i}
\label{e113}
\end{equation}
\begin{equation}
\Gamma^0_{ij} = \Gamma^0_{ji} = - \frac{1}{2c} \left ( \frac{\partial \Xi_i}{\partial x^j} + \frac{\partial \Xi_j}{\partial x^i} + \frac{\partial h_{ij}}{\partial t} \right )
\label{e114}
\end{equation}
\begin{equation}
\Gamma^i_{00} = \frac{1}{c^2} \delta^{ij} \left ( \frac{\partial \Xi_j}  {\partial t}
+ \frac{\partial \Phi}{\partial x^j} \right )
\label{e115}
\end{equation}
\begin{equation}
\Gamma^i_{j0} = \Gamma^i_{0j} = \frac{1}{2c} \delta^{ik} \left ( \frac{\partial \Xi_k}{\partial x^j} - \frac{\partial \Xi_j}{\partial x^k} - \frac{\partial h_{jk}}{\partial t} \right )
\label{e116}
\end{equation}
\begin{equation}
\begin{split}
    \Gamma^i_{jk} \; = \Gamma^i_{kj} = \frac{1}{2} \delta^{il} \left ( \frac{\partial h_{jk}} {\partial x^l} - \frac{\partial h_{jl}} {\partial x^k} - \frac{\partial h_{kl}} {\partial x^j} \right ) \,, 
\label{e117}
\end{split}
\end{equation}
with $h_{ij}$ given by equation (92). For the contracted Christoffel symbol appearing in equation (103) we obtain, considering the linear regime,
\begin{equation}
\begin{split}
\Gamma^\gamma_{\alpha \gamma}
= \frac{1}{2}  \frac{\partial \ln (-g)}{\partial x^\alpha}
\approx \frac{1}{2} \frac{\partial \ln (1 + h)}{\partial x^\alpha} \approx \frac{1}{2} \frac{\partial h}{\partial x^\alpha}
\,. 
\label{e118}
\end{split}
\end{equation}
On the other hand, the 4-dimensional trace $h$ of the metric perturbation is given by
\begin{equation}
h = \eta^{\mu \nu} h_{\mu \nu} 
= \frac{2\Phi}{c^2} - 3\Theta
\,, \label{e119}
\end{equation}
where use was made of equations (23), (90) and (92). Substituting equation (119) in equation (118) we find
\begin{equation}
\begin{split}
\Gamma^\gamma_{\alpha \gamma} = \frac{\partial }{\partial x^\alpha} \left ( \frac{\Phi}{c^2} - \frac{3 \Theta}{2} \right )
\,,
\label{e120}
\end{split}
\end{equation}
which is correct to the first order. It is not very useful to insert all the formulae (112)-(117) and (120) in equations (106) and (107). The result would be cumbersome and difficult to manage. It is much more valuable to use the obtained formulae whenever they are necessary, depending on what dynamical mode of the gravitational field we are dealing. For example, to obtain equations (110) and (111) for plane gravitational waves propagating in the $z$ direction the only non zero Christoffel symbols are those depending on $h^{TT}_{ij}$, and $\Gamma^\gamma_{\alpha \gamma} = 0$, which facilitates calculations.

\section{Non-collisional plasmas: the Einstein-Vlasov-Maxwell system}

A plasma can be defined as a large quasi-neutral collection of charged particles whose dynamics is governed by collective interactions instead of simple pair interactions. Plasma kinetic theory, in turn, can be viewed as an attempt to establish, in a statistical way, the physics involved in the mutual effects between the electromagnetic field and matter in the plasma state. It is natural, in general relativity, to try this statistical approach to study the mutual effects between the gravitational field, the electromagnetic field and the plasma. So, in this Section we will briefly summarize the basic concepts of kinetic theory in curved space-time and write down the fundamental equation governing the dynamics of the one-body distribution function for a non-collisional plasma - the Vlasov equation.

\subsection{The Einstein-Maxwell System}

For the sake of completeness and clarity of the text, before we undergo into kinetic theory, it is instructive to review briefly the Einstein and Maxwell equations in absence of matter - the Einstein-Maxwell system - in the linear regime. For a more detailed treatment, one can consult references \cite{weinberg,mis,web,dirac}. In a material medium, these equations do not constitute a closed system and so can not provide a complete description of any physical system, as we will see.

The right side of the Einstein equations (1) depends upon the distribution of mass and energy of the physical system under consideration. In the same way, the right side of the equations governing the dynamics of the electromagnetic tensor $F_{\mu \nu}$, given by 
\begin{equation}
\frac{1}{\sqrt{-g}} \frac{\partial \left( \sqrt{-g} F^{\mu \nu} \right ) }{\partial x^{\mu}} = \mu_0 J^{\nu}
\,,
\label{e121}
\end{equation}
depends upon the electric 4-current $J^{\nu}$ defined by
\begin{equation}
    J^{\nu} = \left (\rho c , \mathbf{J} \right )
\,,\label{e122}
\end{equation}
where $\rho$ is the electric charge density and $\mathbf{J}$ is the electric current density 3-vector. As usual, $\mu_0$ in equations (121) symbolizes the magnetic permeability of free space, whose value is $4\pi \times 10^{-7}$ H/m. With $\nu = 0$ equations (121) corresponds to Gauss law of electricity, and with $\nu = i$ to the Ampère-Maxwell law. The equations corresponding to the nonexistence of magnetic monopoles (or the Gauss law of magnetism) and to the Faraday-Lenz induction law are given by
\begin{equation}
\frac{\partial F_{\nu\rho}}{\partial x^{\mu}} + \frac{\partial F_{\rho\mu}}{\partial x^{\nu}} + \frac{\partial F_{\mu\nu}}{\partial x^{\rho}} = 0 \,. \label{e123}
\end{equation}
Equations (121) and (123) are the Maxwell equations in curved space-time and, except for the factors $\sqrt{-g}$ in equation (121) are the same as in the flat space-time. Maxwell equations could too be written in terms of the electromagnetic 4-potential $A_{\mu}$, but the forms adopted here involves only the gauge free quantities $F_{\mu \nu}$ given in matrix representation by
\begin{equation}
    \begin{split}
    \mathbb{F} =
        \begin{bmatrix}
            F_{\mu \nu}
        \end{bmatrix} = \begin{bmatrix}
            0 & \frac{E_x}{c} & \frac{E_y}{c} & \frac{E_z}{c}\\
            -\frac{E_x}{c} & 0 & -B_z & B_y \\
            -\frac{E_y}{c} & B_z & 0 & -B_x \\
            -\frac{E_z}{c} & -B_y & B_x & 0 \\
        \end{bmatrix}
    \end{split} \,, \label{e124} 
\end{equation}
that is, the components of the conventional electric and magnetic fields. So, as our aim is a gauge free treatment, the equations (121) and (123) are most adequate for our purposes. Furthermore, in the linear regime equation (121) can be written in the form
\begin{equation}
\frac{\partial F^{\mu \nu}}{\partial x^{\mu}} = \mu_0 J^{\nu} - \frac{1}{2} \frac{\partial h}{\partial x^\mu} F^{\mu \nu}
\,, \label{e125}
\end{equation}
in which the second term at the right side represents an explicit first order influence of the gravitational field over the electromagnetic field. Equation (125) alone provides a nice insight: the direct coupling of electromagnetism to gravity involves only the scalar components of the gravitational field (see equation (119)) and, therefore, there is no direct influence of gravitational wave modes on the dynamics of electromagnetic fields.

To investigate the reciprocal effects, that is, the influence of electromagnetism in the dynamics of the gravitational field, we proceed by conveniently writing the electromagnetic stress-energy tensor in the matrix form  
\begin{equation}
    \mathbb{T}_{em} = \frac{1}{\mu_0} \left [ \mathbb{F} \mathbb{g}^{-1} \mathbb{F} - \frac{1}{4} \mathbb{g} \, \text{Tr} (\mathbb{\Tilde{F} \mathbb{F}} ) \right ] \label{e126}
\end{equation}
where $\mathbb{T}_{em}$, $\mathbb{g}$, $\mathbb{F}$ stands for the matrix representations of the electromagnetic stress-energy tensor, the metric tensor and the electromagnetic tensor in terms of covariant components, and $\mathbb{\Tilde{F}}$ is given by 
\begin{equation}
    \mathbb{\Tilde{F}} =
     \begin{bmatrix}
            F^{\mu \nu}
     \end{bmatrix} = \mathbb{g}^{-1}\mathbb{F} \, \mathbb{g}^{-1} \label{e127}
\end{equation}
where $\mathbb{g}^{-1}$ is the inverse of $\mathbb{g}$. In view of equation (6), to the first order we have
\begin{equation}
    \mathbb{g} = \eta + \mathbb{h} \label{e128}
\end{equation}
and
\begin{equation}
    \mathbb{g}^{-1} = \eta - \mathbb{h} \,, \label{e129}
\end{equation}
where $\eta$ is the matrix representation of the Minkowski metric. So, to the first order we have
\begin{equation}
    \begin{split}
    \mathbb{\Tilde{F}} =
      \begin{bmatrix}
            0 & -\frac{E_x}{c} & -\frac{E_y}{c} & -\frac{E_z}{c}\\
            \frac{E_x}{c} & 0 & -B_z & B_y \\
            \frac{E_y}{c} & B_z & 0 & -B_x \\
            \frac{E_z}{c} & -B_y & B_x & 0 \\
        \end{bmatrix}
    \end{split} \, \label{e130} 
\end{equation}
and
\begin{equation}
    \frac{1}{4\mu_0}\text{Tr} (\mathbb{\Tilde{F} \mathbb{F}} ) = \frac{\epsilon_0 E^2}{2} - \frac{B^2}{2 \mu_0} \,.
    \label{e131}
\end{equation}
For convenience, in equation (131) we introduce the electric permittivity of free space, $\epsilon_0 = \frac{1}{\mu_0 c^2}$. With equations (128), (129) and (131) inserted in equation (126) we obtain, to the first order,
\begin{equation}
    \mathbb{T}_{em} = \mathbb{T}^{(flat)}_{em} + \mathbb{T}^{(curved)}_{em} \,,
    \label{e132}
\end{equation}
where $\mathbb{T}^{(flat)}_{em}$ is the flat space-time contribution to $\mathbb{T}_{em}$, given by
\begin{equation}
    \begin{split}
    \mathbb{T}^{(flat)}_{em} =
    \begin{bmatrix}
    \mathcal{U}^{(flat)}_{em} & -\frac{s_x}{c} & -\frac{s_y}{c} & -\frac{s_z}{c}\\
    -\frac{s_x}{c} & -\tau_{11} & -\tau_{12}  & -\tau_{13} \\
    -\frac{s_y}{c} & -\tau_{21} & -\tau_{22} & -\tau_{23} \\
    -\frac{s_z}{c} & -\tau_{31} & -\tau_{32} & -\tau_{33} \\
    \end{bmatrix}
    \end{split} \,. \label{e133} 
\end{equation}
where $\mathcal{U}^{(flat)}_{em} = \frac{\epsilon_0 E^2}{2} + \frac{B^2}{2\mu_0}$ is the zero order electromagnetic energy density. In equation (133) we employ the definition of the Maxwell stress tensor $\tau_{ij}$
\begin{equation}
   -\tau_{ij} = T^{(flat)}_{em,ij} = -\epsilon_0 E_iE_j -\frac{B_iB_j}{\mu_0} + \mathcal{U}^{(flat)}_{em} \delta_{ij} \,, \label{e134} 
\end{equation}
and indicate the components of the Poynting vector $\mathbf{s} = \frac{1}{\mu_0} \mathbf{E} \times \mathbf{B}$ (the lowercase $\mathbf{s}$ must not to be confused with the capital $\mathbf{S}$ defined in equation (51)). In turn, the first order correction $\mathbb{T}^{(curved)}_{em}$ including the gravitational-electromagnetic coupling is given by
\begin{equation}
    \mathbb{T}^{(curved)}_{em} = -\frac{1}{\mu_0} \left [ \mathbb{F} \mathbb{h} \mathbb{F} + \frac{1}{4} \mathbb{h} \, \text{Tr} (\mathbb{\Tilde{F} \mathbb{F}} ) \right ] \, \label{e135}
\end{equation}
The general form of $\mathbb{T}^{(curved)}_{em}$ is hard to handle, and a better and cleaner job can be made evaluating the first order terms in specific cases and then identify the quantities $T_{00}$ $S$, $S_i$, $P$ and $\sigma_{ij}$, necessary to write Einstein equations (86)-(89). A general procedure to extract these quantities from the stress-energy tensor in reciprocal space will be discussed in part D of the present Section.

As we can see, the Einstein-Maxwell system is a far rich system of equations. On the other hand, it suffers for not taking into account the mutual effects between the gravitational and electromagnetic fields and matter in a material medium, and so can not provide a complete description of the field-matter interactions. Therefore, we will now move on to the kinetic theory as an attempt to describe these interaction and formulate a complete theory of plasmas in curved space-time. Latter, we will apply this theory to the problem of wave propagation in this complex medium.

\subsection{The Vlasov equation in curved space-time}

Plasmas are systems of many electrically charged particles (mostly partially or fully ionized gases). So, plasmas are, at the same time, affected by the gravitational and electromagnetic fields and sources of  these field. It is then necessary to add to the Einstein-Maxwell system one more equation to take into account the effects of the fields on the plasma, an objective that can be achieved through kinetic theory \cite{cerci}.

In analogy to the phase space of non-relativistic mechanics, the phase space of general relativity is also the space of all coordinates and momenta of a physical system. There are, however, some subtleties to consider in the relativistic case. They are:

(i) In theory of relativity time is, naturally, a coordinate, and the construction of the phase space must also incorporate this fact naturally.

(ii) The components of the 3-momentum of a particle and its total (rest plus kinetic) energy are, respectively, the spatial and temporal components of the momentum 4-vector, whose magnitude is constant; that is, there is a constraint between the components of the 4-momentum, which reduces the dimensionality of the phase space.

(iii) Some care is needed in defining the relevant volume elements, in order to preserve the covariance of the theory.

With the precautions mentioned above, we can define the relativistic phase space of an $N$-particle system as the space of the $4N$ space-time coordinates (say, $x$, $y$, $z$ and $x^0 = ct$ of each particle) and the $3N$ spatial components of the 4-moments of the system (say, $p_x$, $p_y$ and $p_z$ of each particle) of the system (the reason why we are apparently neglecting the temporal components of the 4-momenta is the constraint mentioned in item (ii) and will be discussed shortly). It is therefore a $7N$-dimensional space. We will see that, despite the difference in dimensionality between relativistic and non-relativistic phase spaces (the first is $7N$-dimensional and the second is $6N$-dimensional), the one-body distribution function here also depends on 7 variables: in relativistic theory, $x^{\mu}$ and $p^i$ take the place of \textbf{x}, $t$ and \textbf{p}. From a geometric point of view, the $7N$-dimensional phase space can be visualized as a kind of union between space-time (a 4-dimensional manifold $\mathcal{M}$ in which the dynamics of all $N$ particles develop) and the set of tangent spaces at every point of $\mathcal{M}$ (in which the 4-momentum of the $N$ particles resides). Usually, the tangent space of $\mathcal{M}$ at point $P$ is designated by $T_P\mathcal{M}$ and the set of all tangent spaces of $\mathcal{M}$ is called the tangent bundle. However, as a result of the constraint
\begin{equation}
p^\mu p_\mu = p^0 p_0 + g_{ij}p^ip^j = m^2 c^2,  \label{e136}
\end{equation}
we can express one of the 4-momentum components in terms of the others, and the most natural choice (i.e., the one that makes everything more like the usual non-relativistic theory) is to express the temporal component as a function of the spatial components. Thus, the “slice” of the tangent bundle that interests us in the construction of the phase space is the one in which equation (136) is satisfied for each particle of the system. Observe that we are symbolizing the contravariant components of the 4-momentum and the 3-momentum equally by $p^i$, as they are identical.  On the other hand, writing $p_i$ we are indicating only the covariant components of the 3-momentum.
\begin{figure}[!hbt]
\begin{center}
\includegraphics[width=8.0cm,height=6.0cm]{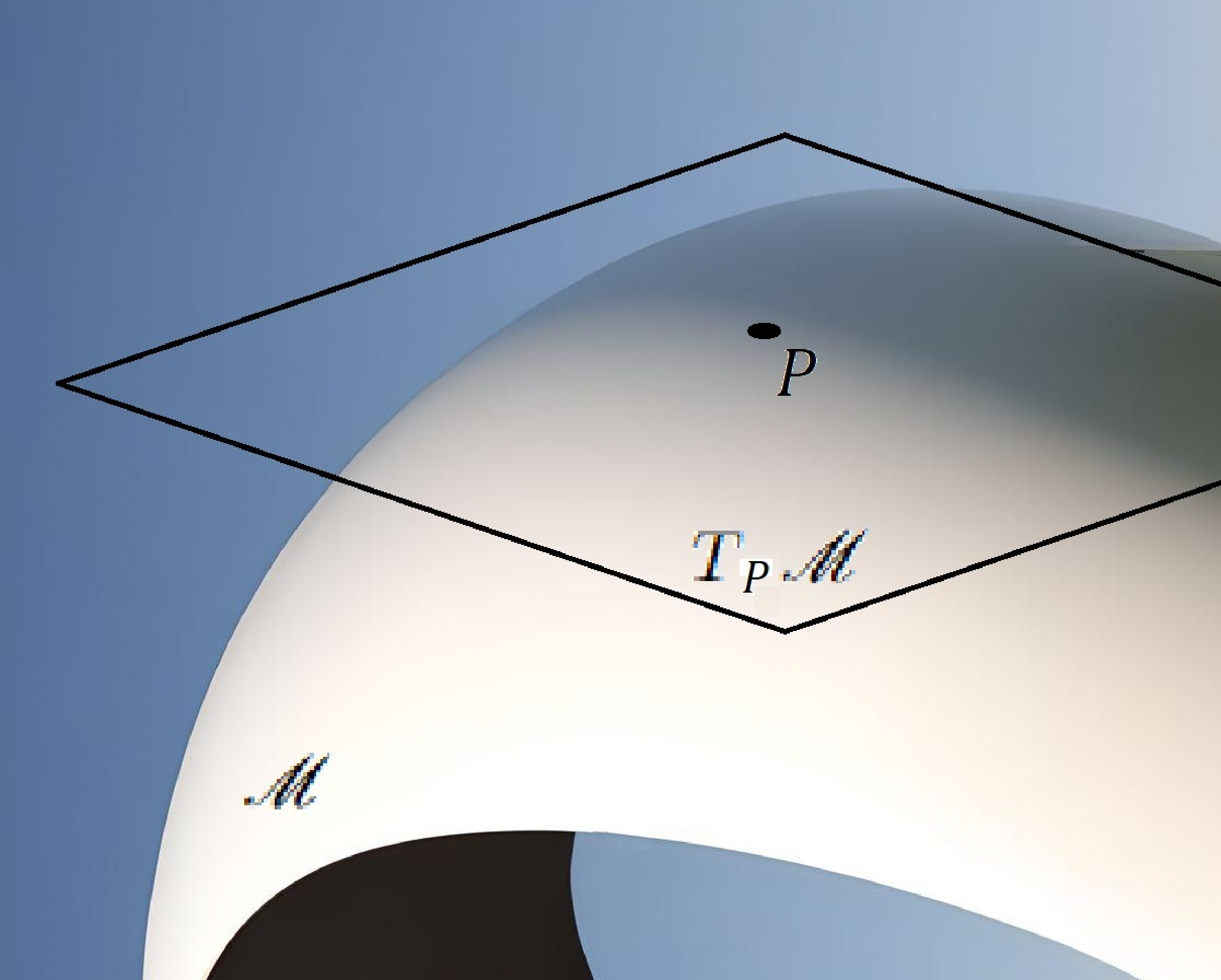}
\end{center}
\caption{The phase space of general relativistic kinetic theory is made up of space-time (a 4-dimensional manifold) and the tangent bundle (the set of all tangent spaces).}
\label{figure1}
\end{figure}

Having defined the relevant phase space, we can move on to the determination of the Vlasov equation in curved space-time, that is, the differential equation governing the dynamics of the one-body distribution function for a collisionless plasma in general relativity. For brevity, as the details of this subject can be found and are cleanly discussed elsewhere \cite{cerci}, we will present the mentioned equation by simple analogy to its non-relativistic counterpart, much in the same way Maxwell equations can be adapted to general relativity. Thus, by imposing the covariance requirement, we write down relativistic equations that fall into the non-relativistic equations in appropriate limits. The non-relativistic Vlasov equation can be written in the form
\begin{equation}
    \frac{df}{dt} = \frac{\partial f}{\partial t} + \frac{dx^i}{dt}\frac{\partial f}{\partial x^i} + \frac{dp^i}{dt}\frac{\partial f}{\partial p^i} = 0 \,, \label{e137}  
\end{equation}
where $f=f(\mathbf{x},\mathbf{p},t)$ is the one-body distribution function. In general relativity, the dependencies of $f$ must be replaced by $x^{\mu}$ and $p^i$. So, the relativistic Vlasov equation can be readily written in the form
\begin{equation}
    \frac{df}{d\tau} = 
    \frac{d x^{\mu}}{d\tau}\frac{\partial f}{\partial x^{\mu}} + \frac{d p^i}{d\tau}\frac{\partial f}{\partial p^i}  = 0 \,, \label{e138}    
\end{equation}
where $f = f(x^{\mu},p^i) \equiv f(x,p)$. On the other hand, from equation (5) and using that $p^{\mu} = m \frac{d x^{\mu}}{d \tau}$, we obtain
\begin{equation}
\frac{d p^i}{d\tau} = -\Gamma_{\nu\rho}^i \, p^{\nu} \frac{d x^{\rho}}{d\tau} + q {F^i}_{\nu} \frac{d x^{\nu}}{d\tau} \,. \label{e139}
\end{equation}%
Finally, substituting equation (139) in equation (138) we find the Vlasov equation in the form
\begin{equation}
    p^{\mu} \frac{\partial f}{\partial x^{\mu}} -\Gamma_{\nu\rho}^i \, p^{\nu} p^{\rho}\frac{\partial f}{\partial p^i} + q {F^i}_{\nu} p^{\nu}\frac{\partial f}{\partial p^i}  = 0 \,. \label{e140}    
\end{equation}
The weak field limit of the Vlasov equation will be discussed in subsection E of the present Section. However, it is interesting to keep in mind right away that the Christoffel symbols (which personify the gravitational force field), in our preferred gauge can all be expressed solely in terms of the gravitational gauge invariants, as pointed out in Section II. It is also nice to check that the non-relativistic Vlasov equation (137) is readily recovered from its general relativistic counterpart, equation (140), by taking the appropriate limits.

The Vlasov equation determines the influence of gravity and electromagnetism over the distribution function - that is, over the matter behavior - via the Christoffel symbols $\Gamma_{\nu\rho}^\mu$ and the electromagnetic tensor ${F^\mu}_\nu$, being the third cornerstone in the theory discussed. In the next subsection we will see how to properly formulate expressions for the matter stress-energy tensor and charge 4-current in terms of the distribution function $f$. For now, it is worth to mention that for a system with many particle species, there is one distribution function for each of these species, each described by an appropriate Vlasov equation.

\subsection{Charge 4-current and matter stress-energy tensor in kinetic theory}

In non-relativistic kinetic theory we can obtain several interesting physical quantities by performing integrations of the one-body distribution function in momentum space. In particular, the number density, the charge density, the charge current density and the matter stress tensor are given, respectively, by
\begin{equation}
    n(\mathbf{x}, t) = \int f(\mathbf{x}, \mathbf{p}, t)\, d^3p \, \label{e141}
\end{equation}
\begin{equation}
    \rho (\mathbf{x}, t) = q \int f(\mathbf{x}, \mathbf{p}, t)\, d^3p \, \label{e142}
\end{equation}
\begin{equation}
    J^i(\mathbf{x}, t) = \frac{q}{m} \int p^i f(\mathbf{x}, \mathbf{p}, t)\, d^3p \, \label{e143}
\end{equation}
\begin{equation}
    T^{ij}(\mathbf{x}, t) = \frac{1}{m} \int p^i \, p^j f(\mathbf{x}, \mathbf{p}, t)\, d^3p \,. \label{e144}
\end{equation}
For the general relativistic counterparts of equations (141)-(144) above, it can be shown \cite{cerci} that we must replace the momentum space volume element $d^3p$ by the invariant $\frac{\sqrt{-g}}{p_0}d^3p$, taking care with the other factors in order to maintain the dimensional consistency of the formalism. Furthermore, the charge density and current combine to form the 4-vector $J^{\mu}$, and the stress tensor is replaced by the most general object $T^{\mu \nu}$. It gives us
\begin{equation}
    n(x) = mc \int f(x, p)\, \frac{\sqrt{-g}}{p_0} d^3p \, \label{e145}
\end{equation}
\begin{equation}
    J^{\mu}(x) = qc \int p^{\mu} f(x, p)\, \frac{\sqrt{-g}}{p_0} d^3p \, \label{e146}
\end{equation}
\begin{equation}
    T^{\mu \nu}(x) = c \int p^{\mu} p^{\nu} f(x, p)\, \frac{\sqrt{-g}}{p_0} d^3p \,. \label{e147}
\end{equation}
It should be noted that, in view of equation (136), in all the equations above we have
\begin{equation}
    p_0 = \sqrt{g_{00} m^2c^2 + (g_{0i}g_{0j} - g_{00}g_{ij})p^ip^j} \,, \label{e148}
\end{equation}
with the contravariant component of the temporal part of the 4-momentum given by 
\begin{equation}
    p^0 = \frac{p_0 - g_{0i}p^i}{g_{00}} \,. \label{e149}
\end{equation}
Our next task is to establish equations (146) and (147) in the weak field limit. For this, we must first obtain the linearized versions of $\frac{\sqrt{-g}}{p_0}$, $p^0 \frac{\sqrt{-g}}{p_0}$ and $p^0 p^0 \frac{\sqrt{-g}}{p_0}$. From equations (148) and (149), we find
\begin{equation}
    \frac{\sqrt{-g}}{p_0} = \frac{1 +  \alpha(x,p)}{\Bar{p}_0}
    \,, \label{e150}
\end{equation}
\begin{equation}
    p^0 \frac{\sqrt{-g}}{p_0} = 1 + \beta(x,p)
    \,, \label{e151}
\end{equation}
\begin{equation}
    p^0 p^0 \frac{\sqrt{-g}}{p_0} = \Bar{p}_0 \left ( 1 + \gamma(x,p) \right )
    \,. \label{e152}
\end{equation}
where $\Bar{p}_0=\sqrt{m^2c^2 + \delta_{ij}p^ip^j} $ is the flat space-time covariant time component of the 4-momentum and $\alpha$, $\beta$ and $\gamma$ are the following non-dimensional first order quantities
\begin{equation}
    \alpha (x,p) = \frac{h}{2} - 
    \frac{h_{00}(\Bar{p}_0)^2 - h_{ij}p^ip^j}{2 (\Bar{p}_0)^2}
    \,, \label{e153}
\end{equation}
\begin{equation}
    \beta(x,p) = \frac{h}{2} - 
    \frac{h_{00} \Bar{p}_0 + h_{0i}p^i}{\Bar{p}_0}
    \,, \label{e154}
\end{equation}
\begin{equation}
    \gamma (x,p) = -\alpha(x,p) + 2\beta(x,p)
    \,. \label{e155}
\end{equation}
Here too, the general expressions of the above equations in terms of the gauge invariants are very complicated and difficult to handle, making it more practical to obtain explicit expressions only in specific calculations. For example, for gravitational waves, we have $h_{00} = h_{0i} = h = 0 $ and  $h_{ij} = h^{TT}_{ij}$. Thus, in this case,
\begin{equation}
    \alpha (x,p) = -\gamma (x,p) = \frac{h^{TT}_{ij}p^ip^j}{2(\Bar{p}_0)^2}
    \, \label{e156}
\end{equation}
\begin{equation}
    \beta (x,p) = 0
    \,. \label{e157}
\end{equation}
As another example, consider the Schwarzschild-like metric in isotropic coordinates for points far away from the event horizon, given by $h_{00} = \frac{2 \Phi}{c^2}$, $h_{ij} = \delta_{ij} \frac{2 \Phi}{c^2}$, $h = -\frac{4 \Phi}{c^2}$ and $h^{TT}_{ij} = h_{0i} = 0$. In this case we have
\begin{equation}
    \alpha (x,p) = \frac{\Phi}{c^2} \left [\frac{\delta_{ij}p^ip^j}{(\Bar{p}_0)^2} - 3 \right ]
    \, \label{e158}
\end{equation}
\begin{equation}
    \beta (x,p) = -\frac{4\Phi}{c^2}
    \, \label{e159}
\end{equation}
\begin{equation}
    \gamma (x,p) = -\frac{\Phi}{c^2} \left [\frac{\delta_{ij}p^ip^j}{(\Bar{p}_0)^2} + 5 \right ]
    \,. \label{e160}
\end{equation}
With formulae (150)-(152), in weak field limit the components of the charge 4-current and of the stress-energy tensor are expressed as
\begin{equation}
   J^i(\mathbf{x}, t) = q \,c \int p^i \frac{\left [1+\alpha(x,p) \right ]}{\Bar{p}_0} f(x,p) d^3p \, \label{e161}
\end{equation}
\begin{equation}
   \rho (\mathbf{x}, t) = \frac{J^0(\mathbf{x}, t)}{c} = q \int \left [1+\beta(x,p) \right ] f(x,p) d^3p \, \label{e162}
\end{equation}
\begin{equation}
   T^{ij}(\mathbf{x}, t) = c \int p^i p^j \frac{\left [1+\alpha(x, p) \right ]}{\Bar{p}_0} f(x,p) d^3p \, \label{e163}
\end{equation}
\begin{equation}
   T^{0i} (\mathbf{x}, t) = c \int p^i \left [1+\beta(x,p) \right ] f(x,p) d^3p \, \label{e164}
\end{equation}
\begin{equation}
   T^{00} (\mathbf{x}, t) = c \int \Bar{p}_0 \left [1+\gamma(x,p) \right ] f(x,p) d^3p \, \label{e165}
\end{equation}
Equations (161) and (162) represents the source terms for Maxwell equations, whereas equations (163)-(165) play that role in Einstein equations. In the next subsection we will see how to extract from the components $T^{0i}$ and $T^{ij}$ the objects $S$, $S_i$, $P$ and $\sigma_{ij}$ (necessary to write Einstein equations), a job that proves to be simpler in the reciprocal space.

\subsection{Fourier transformed source terms for gravity}

In order to extract $S$, $S_i$, $P$ and $\sigma_{ij}$ from $T^{0i}$ and $T^{ij}$, it is convenient to take the Fourier transformed versions of equations (51) and (52) for $T^{0i}$ and (70)-(73) for  $T^{ij}$. Furthermore, taking Fourier transforms is the usual starting point for the study of waves in a material media, providing a simple way to obtain dispersion relations, as will be done in Section IV.  We employ the convention in which the Fourier transform of a function $f(\mathbf{x})$  and the related Fourier integral are respectively given by
\begin{equation}
    \hat{f}(\mathbf{k}) = \int f(\mathbf{x}) e^{-i\mathbf{k} \cdot \mathbf{x}} d^3x \, \label{e166}
\end{equation}
and
\begin{equation}
    f(\mathbf{x}) = \frac{1}{(2\pi)^{3}} \int \hat{f}(\mathbf{k}) e^{i\mathbf{k} \cdot \mathbf{x}} d^3k \,. \label{e167}
\end{equation}
Hereafter, a hat over any letter will symbolize a spatial Fourier transform, as above. By Fourier transforming equations (51),  (52) and (70)-(73) we obtain
\begin{equation}
\hat{T}_{0i} = c \left (ik_i \hat{S} + \hat{S}_i \right ) \;, \label{e168}
\end{equation}
\begin{equation}
k^i \hat{S}_i = 0 \;, \label{e169}
\end{equation}
\begin{equation}
\begin{split}
\hat{T}_{ij} = & \; \hat{\sigma}_{ij} + \delta_{ij}\hat{P}
+ i\left ( k_j\hat{\sigma}_i + k_i\hat{\sigma}_j \right ) \\
& + \left ( k_i k_j - \frac{1}{3} \delta_{ij} k^2 \right ) \hat{\sigma}
\;, \label{e70}
\end{split}
\end{equation}

\begin{equation}
k^i \hat{\sigma}_{ij} = 0 \;, \label{e171}
\end{equation}
\begin{equation}
\delta^{ij} \hat{\sigma}_{ij} = 0 \;, \label{e172}
\end{equation}
\begin{equation}
k^i \hat{\sigma}_i = 0 \;, \label{e173}
\end{equation}
where $k^2 = k^ik_i = \delta_{ij}k^ik^j$ (remember that $\mathbf{k}$ is a 3-vector, not a 4-vector). Contracting equation (168) with $k^i$ and taking into account equation (169) we find
\begin{equation}
\hat{S} = \frac{k^i}{ick^2} \hat{T}_{0i} \;, \label{e174}
\end{equation}
Returning with the result (174) in (168) and rearranging, we get
\begin{equation}
\hat{S}_i = \frac{1}{c} \left (\delta^j_i - \frac{k_i k^j}{k^2} \right ) \hat{T}_{0j} \;. \label{e175}
\end{equation}
By imposing conditions (171)-(173) we can also solve equation (170) for $\hat{P}$, $\hat{\sigma}$, $\hat{\sigma}_i$ and $\hat{\sigma}_{ij}$. The results are
\begin{equation}
\hat{P}= \frac{1}{3} \delta^{ij} \hat{T}_{ij} \;, \label{e176}
\end{equation}
\begin{equation}
\hat{\sigma}= -\frac{3}{2} \left ( \frac{\delta^{ij}}{3k^2} - \frac{k^ik^j}{k^4}\right ) \hat{T}_{ij} \;, \label{e177}
\end{equation}
\begin{equation}
\hat{\sigma}_i = \frac{1}{ik^2} \left ( k^k \delta^j_i - \frac{k_ik^jk^k}{k^2}  \right ) \hat{T}_{jk} \;, \label{e178}
\end{equation}
\begin{equation}
\begin{split}
\hat{\sigma}_{ij} = \left (\delta^k_i\delta^l_j - \frac{1}{2}\delta_{ij}\delta^{kl} \right ) \hat{T}_{kl} \;\;\;\;\;\;\;\;\;\;\;\;\;\;\;\;\;\; \\
+ \frac{1}{2k^2}\left ( k_i k_j \delta^{kl} + \delta_{ij} k^k k^l - 2k_j k^k \delta^l_i -2k_i k^k \delta^l_j \right )\hat{T}_{kl} \\ + \frac{1}{2k^4} k_i k_j k^k k^l \hat{T}_{kl} \,. \;\;\;\;\;\;\;\;\;\;\;\;\;
\; \label{e179}
\end{split}
\end{equation}
Although  $\sigma$ and $\sigma_i$  do not appears in the equations for the gravitational field, obtaining equations (177) and (178) for these quantities was important for writing equation (179) for $\sigma_{ij}$. We now have at hand all the expressions we need to write the field equations (86)-(89).

\subsection{3-dimensional forms of Maxwell and Vlasov equations in the weak field regime. The EVM system}

From equations (123), (124), (125) and (130), to the first order in the gravitational perturbation, Maxwell equations can be written in the familiar 3-dimensional form
\begin{equation}
    \nabla \cdot \mathbf{E} = \frac{\rho}{\epsilon_0} -\frac{1}{2}\nabla h \cdot \mathbf{E} \; \label{e184}
\end{equation}
\begin{equation}
\begin{split}
    \nabla \times \mathbf{B} = \mu_0 \mathbf{J} + \frac{1}{c^2} \frac{\partial \mathbf{E}}{\partial t} + \frac{1}{2} \left ( \frac{1}{c^2} \frac{\partial h}{\partial t} \mathbf{E} - \nabla h \times \mathbf{B} \right ) \; \label{e181}
\end{split}
\end{equation}
\begin{equation}
    \nabla \times \mathbf{E} = -\frac{\partial \mathbf{B}}{\partial t} \; \label{e182}
    \end{equation}
\begin{equation}
    \nabla \cdot \mathbf{B} = 0 \; \label{e183}
\end{equation}
where $\nabla = \left ( \frac{\partial}{\partial x} , \frac{\partial}{\partial y} , \frac{\partial}{\partial z}\right )$, as in flat space-time. Equations (180)-(183) corresponds to the well known equations of the electromagnetic theory, corrected by a few terms proportional to space-time derivatives of $h$ and to the electric and magnetic fields themselves. In this same 3-dimensional formalism, Vlasov equation (140) can be written in the following friendly form
\begin{equation}
\begin{split}
\frac{\partial f}{\partial t} + \frac{c \mathbf{p}}{p^0} \cdot \nabla f + \left (\mathbf{F}_{em} + \mathbf{F}_g \right ) \cdot \nabla_{\mathbf{p}} f = 0
\end{split} \; , \label{e184}
\end{equation}
where
\begin{equation}
\begin{split}
\mathbf{F}_{em} = q \left (\mathbf{E} + \frac{c \mathbf{p}}{p^0} \times \mathbf{B} \right )
\end{split} \; \label{e185}
\end{equation}
is the electromagnetic force and
\begin{equation}
\begin{split}
\mathbf{F}_g = \frac{p^0}{c} \left [-\nabla \Phi -\frac{\partial \mathbf{\Xi}}{\partial t} +  \frac{c \mathbf{p}}{p^0} \times \left (\nabla \times \mathbf{\Xi} \right ) \right ] + \mathbf{F}_T
\end{split} \; \label{e186}
\end{equation}
is the gravitational force, both correct to the first order. $\mathbf{F}_T$ comes from the tensor part of the metric and has components given by
\begin{equation}
\begin{split}
F_{T,i} = \frac{\partial h_{ij}}{\partial t} p^j + \frac{c}{p^0} \left [jk,i \right] p^j p^k
\end{split} \; , \label{e187}
\end{equation}
with
\begin{equation}
\begin{split}
\left [jk,i \right] =\frac{1}{2} \left (\frac{\partial h_{ij}}{\partial x^k} + \frac{\partial h_{ik}}{\partial x^j} - \frac{\partial h_{jk}}{\partial x^i} \right )
\end{split} \; . \label{e188}
\end{equation}
For consistency, in the whole set of equations (184)-(187), $p^0$ must be taken correct to the zero order. From equations (148) and (149) we obtain for $p^0$ in this limit
\begin{equation}
\begin{split}
p^0 = \Bar{p}_0 = \sqrt{m^2c^2 + \delta_{ij}p^ip^j} \; . \label{e189}
\end{split}
\end{equation}
In equation (186), the terms collected in brackets closely resembles the electromagnetic force - given by equation (185) - written in terms of electromagnetic potentials (note that just the first one appears in the Newtonian theory of gravitation!). However, it is important to stress that this similarity is somewhat superficial, since the potentials $\Phi$ and $\mathbf{\Xi}$ always satisfy the inhomogeneous Poisson equations (87) and (88) (not first order coupled equations), and there is not a gravitational analogous to the Faraday and Maxwell induction terms in the gravitational field equations (86)-(89).  Furthermore, the gravitational potentials we are dealing are gauge invariant in the linear theory of gravity, whereas electromagnetic potentials are not invariant with respect to the electromagnetic gauge transformations. It is also worth to mention that, while the electromagnetic force is invariant to the gauge transformations of electromagnetism, the gravitational force is dependent on the chosen gravitational gauge, as required by the principle of equivalence.

Einstein equations (86)-(89), Maxwell equations (180)-(183) and Vlasov equation (184) constitutes a complete system to describe the behavior of a collisionless plasma in a general relativistic framework - the EVM system. In the next Section, the dispersion relation for gravitational waves in an homogeneous plasma will be derived from these system of equations.

\section{Gravitational and electrostatic waves in an electron-positron plasma}

So far we have discussed Einstein, Maxwell and Vlasov equations just in the gravitational weak field limit, that is, no approximations were made for the electromagnetic field and for the distribution function. Furthermore, we have yet to introduce in the formalism the fact that plasmas are quasi-neutral systems of (at least) two charged particle species. Hereafter, the whole set of field equations (86)-(89) and (180)-(183) and correspondent source terms, plus the Vlasov equation (184), elaborated in the gravitational weak field limit, will be taken as exact equations. The dispersion function for any kind of oscillation or wave can then be found employing perturbation theory.

\subsection{The Einstein-Vlasov-Poisson system for an homogeneous neutral electron-positron plasma}

Relativistic electron-positron pair plasmas are principal constituents of high-energy astrophysical environments, such as neutron stars and black holes surroundings \cite{che}. Thus, a theoretical description of this type of plasma is fundamental in understanding and interpreting several processes and phenomena that occur in these environments, such as pair creation and annihilation, relativistic jets from active galactic nuclei (AGN), and gamma-ray bursts (GRBs) \cite{che,ruf}. In these systems, collective plasma processes are responsible for determining the magnetic field dynamics, energy partition, and radiation emission \cite{che}. Furthemore, 
this type of pair plasma appeared in the very early universe \cite{rees}.

Although there are several theoretical works on relativistic electron-positron pair plasmas dynamics, to our knowledge there is still a gap regarding an understading the particularities of gravitational oscillatory modes in this medium - especially gravitational waves - employing a kinetic approach. For example, in \cite{roz} a quantum hidrodynamical model for a multicomponent electron-positron-ion plasma was proposed for studying their gravitational instability. In this work, however, gravitation was taken in the Newtonian sense, that is, via the Poisson equation for a scalar potential. Employing a two-fluid model, electromagnetic wave instability in unmagnetized electron-positron pair plasma was discussed in reference \cite{gratt}, and in work \cite{onis} again a two-fluid approach was established for describing nonlinear waves in an inhomogeneous collisionless magnetized relativistic electron-positron plasma in a prescribed gravitational field. The nonlinear interaction between magnetic field-aligned electromagnetic waves and electrostatic oscillation in a electron-positron-ion plasma was considered in reference
\cite{shu}, and in \cite{tom} relativistic collisionless shock waves, associated with GBRs, propagating in inhomogeneous electron-positron plasmas was studied. In these last two works, however, nothing related to gravity was considered.

In view of the above, as an application illustrating the generality, security and simplicity of the gauge invariant formalism, we now apply it to an electron-positron homogeneous plasma where gravitational and electrostatic wave perturbations takes place. First, in order to write down the appropriate equations, we must to establish the relevant physical variables of the problem.  For gravitational waves, the scalar and vector components of the metric perturbation are $\Phi = \Theta = \Xi_i = 0 $. The transversal traceless tensor components are $h^{TT}_{ij} = h_{ij}$ by virtue of equation (92), and so, hereafter we will omit the superscript $TT$ in $h_{ij}$ for simplicity. As pointed out in Section II, part C, the transversality and tracelles conditions for gravitational waves are not restricted to vacuum propagation, being solely based in the Helmholtz decomposition of the metric tensor. So, here too considering a plane gravitational wave propagating along the $z$ axis, in the direction of increasing $z$, we have
\begin{equation}
    \epsilon_{31} = \epsilon_{32} = \epsilon_{33} = 0 
\, \label{e190}
\end{equation}
and
\begin{equation}
    \epsilon_{22} = - \epsilon_{11} 
\,, \label{e191}
\end{equation}
which leads to
\begin{equation}
    \begin{split}
        \begin{bmatrix}
            h_{ij}
        \end{bmatrix}=\begin{bmatrix}
            \epsilon_{11} & \epsilon_{12} & 0\\
            \epsilon_{12} & - \epsilon_{11} & 0\\
            0 & 0 & 0
        \end{bmatrix} e^{i(kz - \omega t)}
    \end{split}  \, \label{e192}
\end{equation}
with $k^1 = k^2 = 0$ and $k^3 = k$. The relevant Einstein equations for the system are equations (89), which here reads as
\begin{equation}
\left( \frac{1}{c^2} \frac{\partial^2}{\partial t^2} - \frac{\partial^2}{\partial z^2} \right) h_{11} = - \frac{16\pi G}{c^4} \sigma_{11} \, \label{e193}
\end{equation}
and
\begin{equation}
\left( \frac{1}{c^2} \frac{\partial^2}{\partial t^2} - \frac{\partial^2}{\partial z^2} \right) h_{12} = - \frac{16\pi G}{c^4} \sigma_{12} \,. \label{e194}
\end{equation}
The tranverse traceless tensor components  $\sigma_{ij}$ are obtained in a straightforwardly manner by employing the general procedure outlined in Section III, part D. From equation (179), with $k^1 = k^2 = 0$ and $k^3 = k$, we find that the only non zero  $\sigma_{ij}$ objects are
\begin{equation}
\sigma_{11} = \frac{1}{2} \left (T_{11} - T_{22} \right )
 \, \label{e195}
\end{equation}
\begin{equation}
\sigma_{22} = - \sigma_{11}
 \, \label{e196}
\end{equation}
\begin{equation}
\sigma_{12} = T_{12}
 \,. \label{e197}
\end{equation}
In turn, the matter stress tensor components $T_{ij}$ are given by equation (163), with the $\alpha$ function given by equation (156). In order to construct these object, we must sum up the positron and the electron contributions. So, symbolizing the positron ant the electron distribution functions respectively by $f$ and $g$, we find
\begin{equation}
   T_{11} = n_0 c \int \frac{(p_1)^2}{{\Bar{p}_0}} \left (1 + \alpha \right ) \left( f + g  \right) d^3p \,, \label{e198}
\end{equation}
\begin{equation}
   T_{22} = n_0 c \int \frac{(p_2)^2}{{\Bar{p}_0}} \left (1 + \alpha \right ) \left( f + g  \right) d^3p \, \label{e199}
\end{equation}
and
\begin{equation}
   T_{12} = n_0 c \int \frac{p_1p_2}{{\Bar{p}_0}} \left (1 + \alpha \right ) \left( f + g  \right) d^3p \,, \label{e200}
\end{equation}
with
\begin{equation}
\alpha = \frac{\left[(p^1)^2 - (p^2)^2 \right ] h_{11} + 2 \, p^1p^2 h_{12}}{2(\Bar{p}_0)^2}
\,. \label{e201}
\end{equation}
In equations (198)-(200) (and in every equation hereafter), as usual in plasma physics, we conveniently renormalize the distribution functions in the way that the momentum space integration of its non relativistic version in an homogeneous equilibrium medium results in the unity (see equation (141)), which, in turn, entails the multiplication of every momentum space integrals of the distribution functions $f$ and $g$ by $n_0$, the equilibrium particle density for positrons and electrons (the same density for both species in a neutral homogeneous plasma). We must not worry about the electromagnetic stress tensor because, as we will see shortly, it is a second order object, thus negligible in the first order approach we begin to discuss below.

Now, as usual in plasma theory, we employ perturbation theory assuming the following first order approximations for the distribution functions and for the metric perturbations:
\begin{equation}
\begin{split}
f(\mathbf{x},\mathbf{p},t) = f^{(0)}(\mathbf{p}) + f^{(1)}(\mathbf{x},\mathbf{p},t)\;, \label{e202}
\end{split}
\end{equation}
\begin{equation}
\begin{split}
g(\mathbf{x},\mathbf{p},t) = g^{(0)}(\mathbf{p}) + g^{(1)}(\mathbf{x},\mathbf{p},t)\;, \label{e203}
\end{split}
\end{equation}
\begin{equation}
\begin{split}
h_{ij}(\mathbf{x},t) = h^{(0)}_{ij} + h^{(1)}_{ij}(\mathbf{x}, t)\;, \label{e204}
\end{split}
\end{equation}
A superscript $(0)$ indicates zero order (or unperturbed equilibrium quantities) and a superscript $(1)$ indicates first order corrections, assumed small. As indicated in equations (202)-(203), the unperturbed distribution functions are assumed to be independent of positions (as the system is homogeneous) and time independent. In view of equations (199)-(202), taking now the unperturbed metric satisfying $h_{ij}^{(0)} = 0$ and assuming that the equilibrium distributions functions for electrons and positrons are equal, that is, $g^{(0)} = f^{(0)}$, in view of equations (195)-(204), Einstein equations (193)-(194) correct to the first order are written as
\begin{equation}
\begin{split}
\left( \frac{1}{c^2} \frac{\partial^2}{\partial t^2} - \frac{\partial^2}{\partial z^2} \right) & h_{11} \\ = - \frac{8\pi G n_0}{c^3} \int \frac{(p_1)^2-(p_2)^2}{{\Bar{p}_0}} & \left (\mathcal{F} + 2 \alpha f^{(0)} \right ) d^3p \, \label{e205}
\end{split}
\end{equation}
and
\begin{equation}
\begin{split}
\left( \frac{1}{c^2} \frac{\partial^2}{\partial t^2} - \frac{\partial^2}{\partial z^2} \right) & h_{12} \\ = - \frac{8\pi G n_0}{c^3} \int \frac{2p_1 p_2}{{\Bar{p}_0}} & \left (\mathcal{F} + 2 \alpha f^{(0)} \right ) d^3p \,, \label{e206}
\end{split}
\end{equation}
where we define a first auxiliary distribution function
\begin{equation}
\mathcal{F} = f^{(1)} + g^{(1)} \, \label {e207}
\end{equation}
and omit the superscript $(1)$ in $h_{ij}$ and $\alpha$ to simplify the notation. Equations (205) and (206), complemented by the formulas (201) and (207), completes our construction of gravitational field equations for the system we are dealing.

We pass now to the elaboration of the relevant Maxwell equations for the electron-positron plasma.  As we are interested in an electrostatic problem, we must assume $\mathbf{B} = 0$. For the electric field, we assume the first order perturbation expansion 
\begin{equation}
\mathbf{E} = \mathbf{E}^{(1)}(\mathbf{x}, t)\; \label{e208}
\end{equation}
with $\mathbf{E}^{(0)} = 0$. So, from equations (180)-(183), with $h = 0$ we are left with
\begin{equation}
\nabla \cdot \mathbf{E} = \frac{\rho}{\epsilon_0}  \; \label{e209}
\end{equation}
and
\begin{equation}
    \nabla \times \mathbf{E} = 0 \; , \label{e210}
\end{equation}
where we again omit the superscript $(1)$ in $\mathbf{E}$ and $\rho$, for simplicity. The equations for $\mathbf{B}$ are not relevant here. Observe that, coherent with the assumption $\mathbf{E}^{(0)} = 0$, the zero order charge density vanishing is guaranteed by the condition of equal equilibrium densities for positrons and electrons and by the assumption $g^{(0)} = f^{(0)}$. With $h = \beta =0$ (see equation (157)) the first order charge density is given by
\begin{equation}
   \rho = \rho_{\text{positrons}} + \rho_{\text{electrons}} = en_0 \int \mathcal{G} d^3p \,, \label{e211}
\end{equation}
where was defined a second auxiliary distribution function
\begin{equation}
\mathcal{G} = f^{(1)} - g^{(1)} \,. \label {e212}
\end{equation}
As is well known, equation (210) ensures that the electric field can be expressed as the gradient of a scalar potential $\Psi$ as
\begin{equation}
\mathbf{E} = - \nabla \Psi \;.  \label{e213}
\end{equation}
Inserting equations (211) and (213) in equation (209), we are led to the Poisson equation
\begin{equation}
\nabla^2 \Psi = -\frac{n_0 e}{\epsilon_0} \int \mathcal{G} d^3p \;  \label{e214}
\end{equation}
for the first order potential $\Psi$. As mentioned before, as the electric field $\mathbf{E}$ and the metric perturbations $h_{ij}$ are first order quantities, equations (132)-(135) shows that, indeed, the electromagnetic stress tensor is a second order object, thus being discarded in the construction of the Einstein equations (205)-(206).

Concluding the construction of the Einstein-Vlasov-Poisson system, we pass now to the relevant Vlasov equations. Following the usual steps of perturbation theory, to the first order the Vlasov equations for positrons and electrons are respectively given by
\begin{equation}
\begin{split}
\frac{\partial f^{(1)}}{\partial t} + \frac{c \mathbf{p}}{\Bar{p}_0} \cdot \nabla f^{(1)} + \left( \mathbf{F}_g - e \nabla \Psi \right) \cdot \nabla_{\mathbf{p}} f^{(0)} = 0  \; \label{e215}
\end{split}
\end{equation}
and
\begin{equation}
\begin{split}
\frac{\partial g^{(1)}}{\partial t} + \frac{c \mathbf{p}}{\Bar{p}_0} \cdot \nabla g^{(1)} + \left( \mathbf{F}_g + e \nabla \Psi \right) \cdot \nabla_{\mathbf{p}} f^{(0)} = 0  \;, \label{e216}
\end{split}
\end{equation}
where use was made of equation (213). We stress that, as the difference of electrons and positrons are just the sign of its charges, the gravitational force is the same for the two types of particles. In view of equations (205)-(206) and (214), it is convenient to rewrite the Vlasov equations (215) and (216) for the auxiliary distributions $\mathcal{F}$ and $\mathcal{G}$ defined by equations (207) and (212). Thus, adding and subtracting equation (216) from equation (215) we get 
\begin{equation}
\begin{split}
\frac{\partial \mathcal{F}}{\partial t} + \frac{c \mathbf{p}}{\Bar{p}_0} \cdot \nabla \mathcal{F} + 2 \, \mathbf{F}_g  \cdot \nabla_{\mathbf{p}} f^{(0)} = 0  \; \label{e217}
\end{split}
\end{equation}
and
\begin{equation}
\begin{split}
\frac{\partial \mathcal{G}}{\partial t} + \frac{c \mathbf{p}}{\Bar{p}_0} \cdot \nabla \mathcal{G} -2 e \, \nabla \Psi \cdot \nabla_{\mathbf{p}} f^{(0)} = 0  \;, \label{e218}
\end{split}
\end{equation}
completing the system of equations.

Note that, to the first order, electricity and gravity are completely decoupled in the electron-positron plasma. First, there is one pair of equations (namely, equations (214) and (218)) to describe electrostatic oscillations in the plasma. Second, there are three equations (equations (205)-(206) and (217)) doing the same job for gravitational waves. In the former case the potential function whose oscillations are considered is $\Psi$ and the relevant distribution is the auxiliary function $\mathcal{G}$. In the last one, the potential functions are $h_{11}$ and $h_{12}$ and the relevant distribution is the auxiliary function $\mathcal{F}$. The problem of electrostatic oscillations thus reduces to the special relativistic case, which is discussed in details elsewhere \cite{laing} and so will not be carried forward here. In the subsection C of this Section, we will just deal with the problem of propagation of gravitational waves in the electron-positron plasma. However, before we undergo in this subject, to complete the picture, it is important to write down expressions for the components of the gravitational force.
\\
\subsection{The force exerted by gravitational waves}

From equation (186) it is clear that, for the gravitational oscillations we are dealing, the only non zero terms of the gravitational force are that related to the tensor part of the metric, given by equation (187) (observe that this components of the gravitational force are just those that have not a Newtonian counterpart and electromagnetic analogues). To find the desired force components, we must first compute the Christoffel symbols of the first kind, given by equation (188). From the twenty seven objects $[jk,i]$, the only non zero are
\begin{equation}
\begin{split}
[1\;3,1] = [3\;1,1] = \frac{1}{2} \frac{\partial h_{11}}{\partial z} 
\; \label{e219}
\end{split}
\end{equation}
\begin{equation}
\begin{split}
[2\;3,1] = [3\;2,1] = \frac{1}{2} \frac{\partial h_{12}}{\partial z} 
\; \label{e220}
\end{split}
\end{equation}
\begin{equation}
\begin{split}
[1\;3,2] = [3\;1,2] = \frac{1}{2} \frac{\partial h_{12}}{\partial z} 
\; \label{e221}
\end{split}
\end{equation}
\begin{equation}
\begin{split}
[2\;3,2] = [3\;2,2] = -\frac{1}{2} \frac{\partial h_{11}}{\partial z} 
\; \label{e222}
\end{split}
\end{equation}
\begin{equation}
\begin{split}
[1\;2,3] = [2\;1,3] = -\frac{1}{2} \frac{\partial h_{12}}{\partial z} 
\; \label{e223}
\end{split}
\end{equation}
\begin{equation}
\begin{split}
[1\;1,3] = -\frac{1}{2} \frac{\partial h_{11}}{\partial z} 
\; \label{e224}
\end{split}
\end{equation}
\begin{equation}
\begin{split}
[2\;2,3] = \frac{1}{2} \frac{\partial h_{11}}{\partial z} 
\;. \label{e225}
\end{split}
\end{equation}
With equations (219)-(225) inserted in equation (187) we find for the gravitational force components the expressions
\begin{equation}
\begin{split}
F_{g,1} = p^1 \Lambda h_{11} + p^2 \Lambda h_{12}
\; \label{e226}
\end{split}
\end{equation}
\begin{equation}
\begin{split}
F_{g,2} = -p^2 \Lambda h_{11} + p^1 \Lambda h_{12}
\; \label{e227}
\end{split}
\end{equation}
\begin{equation}
\begin{split}
F_{g,3} = -\frac{c}{2\Bar{p}_0} \left [(p^1)^2 - (p^2)^2 \right] \frac{\partial h_{11}}{\partial z} - \frac{cp^1p^2}{\Bar{p}_0} \frac{\partial h_{12}}{\partial z}
\; \label{e228}
\end{split}
\end{equation}
where $\Lambda$ is the differential operator
\begin{equation}
\begin{split}
\Lambda = \frac{cp^3}{\Bar{p}_0} \frac{\partial}{\partial z} + \frac{\partial}{\partial t}
\;. \label{e229}
\end{split}
\end{equation}
We are now in position to pursue the dispersion relation of gravitational waves in electron-positron plasma.

\subsection{Dispersion relation for gravitational waves}

To find the dispersion relation for gravitational waves in the studied medium, we could proceed as in the non relativistic theory by taking the Fourier-Laplace transform (that is, the Fourier transform in space and Laplace transform in time) of the Vlasov and field equations, properly treating the problem as an initial value one. Alternatively, a simpler although equivalent procedure is to Fourier transform the equations, allowing a complex angular frequency with a (presumably) small imaginary part. So, employing the usual prescriptions
\begin{equation}
\begin{split}
\frac{\partial}{\partial t} \rightarrow -i \omega
\; \label{e230}
\end{split}
\end{equation}
and
\begin{equation}
\begin{split}
\frac{\partial}{\partial z} \rightarrow i k
\; \label{e231}
\end{split}
\end{equation}
applying the second method to Einstein equations (205)-(206) and to Vlasov equation (217), we are led to the following set of transformed equations:
 \begin{equation}
\begin{split}
& \left( \omega^2 - c^2k^2 \right) \Tilde{h}_{11} \\ & = \frac{8\pi G n_0}{c} \int \frac{(p_1)^2-(p_2)^2}{{\Bar{p}_0}} \left (\Tilde{\mathcal{F}} + 2 \Tilde{\alpha} f^{(0)} \right ) d^3p \, \label{e232}
\end{split}
\end{equation}
and
\begin{equation}
\begin{split}
& \left( \omega^2 - c^2k^2 \right) \Tilde{h}_{12} \\ & = \frac{8\pi G n_0}{c} \int \frac{2p_1 p_2}{{\Bar{p}_0}} \left (\Tilde{\mathcal{F}} + 2 \Tilde{\alpha} f^{(0)} \right ) d^3p \, \label{e233}
\end{split}
\end{equation}

\begin{equation}
\begin{split}
\Tilde{\Lambda} \Tilde{\mathcal{F}} = -2 \, \Tilde{\mathbf{F}}_g  \cdot \nabla_{\mathbf{p}} f^{(0)}  \; \label{e234}
\end{split}
\end{equation}
Furthermore, by Fourier transforming the $\alpha $ function (equation (201)) and the gravitational force components (equations (226)-(228) we find
\begin{equation}
\Tilde{\alpha} = \frac{\left[(p^1)^2 - (p^2)^2 \right ] \Tilde{h}_{11} + 2 \, p^1p^2 \Tilde{h}_{12}}{2(\Bar{p}_0)^2}
\,. \label{e235}
\end{equation}
\begin{equation}
\begin{split}
\Tilde{F}_{g,1} = p^1 \Tilde{\Lambda} \Tilde{h}_{11} + p^2 \Tilde{\Lambda} \Tilde{h}_{12}
\; \label{e236}
\end{split}
\end{equation}
\begin{equation}
\begin{split}
\Tilde{F}_{g,2} = -p^2 \Tilde{\Lambda} \Tilde{h}_{11} + p^1 \Tilde{\Lambda} \Tilde{h}_{12}
\; \label{e237}
\end{split}
\end{equation}
\begin{equation}
\begin{split}
\Tilde{F}_{g,3} = -i \Bar{p}_0 c \, k  \Tilde{\alpha}
\;. \label{e238}
\end{split}
\end{equation}
In the set of equations above it was defined
\begin{equation}
\begin{split}
\Tilde{\Lambda} = i \left(\frac{cp^3k}{\Bar{p}_0} - \omega \right) \; \label{e239}
\end{split}
\end{equation}
and a tilde over any letter symbolizes a space-time Fourier transform, defined as
\begin{equation}
    \Tilde{f}(\mathbf{k}, \omega) = \int f(\mathbf{x}, t) e^{-i \left( \mathbf{k} \cdot \mathbf{x} - \omega t \right ) } dt \, d^3x \, \label{e240}
\end{equation}
for a function $f$. The correspondent Fourier integral (or inverse Fourier transform) is, therefore, given by
\begin{equation}
    f(\mathbf{x}, t) = \frac{1}{(2\pi)^{4}} \int \hat{f}(\mathbf{k}, \omega) e^{i \left( \mathbf{k} \cdot \mathbf{x} - \omega t \right ) } d\omega \, d^3k \,. \label{e241}
\end{equation}
Just to mention, in view of equation (241) and the form (192) assumed for the gravitational waves, we identify the components of the gravitational polarization tensor as
\begin{equation}
\epsilon_{ij} = \frac{1}{(2\pi)^{4}} \Tilde{h}_{ij} 
\,. \label{e242}
\end{equation}
Our efforts now relies in writing the functions $\Tilde{\alpha}$, $\Tilde{\mathcal{F}}$ and $\Tilde{F}_{g,i}$  in terms of $\Tilde{h}_{11}$ and $\Tilde{h}_{12}$ and then to substitute the results in the field equations (232) and (233). From equations (234)-(239), we obtain
\begin{equation}
\begin{split}
& \Tilde{\mathcal{F}} + 2 \Tilde{\alpha} f^{(0)} \\ 
& = 2 \left [ \frac{(p^1)^2-(p^2)^2}{2 (\Bar{p}_0)^2} - \left(p^1 \frac{\partial}{\partial p^1} - p^2 \frac{\partial}{\partial p^2} \right) \right ] f^{(0)} \Tilde{h}_{11} \\
& + 2 \left [ \frac{p^1p^2}{(\Bar{p}_0)^2} - \left(p^2 \frac{\partial}{\partial p^1} + p^1 \frac{\partial}{\partial p^2} \right) \right ] f^{(0)} \Tilde{h}_{12} \\
& + 2 \frac{ck}{ \left ( \frac{c \, p^3}{\Bar{p}_0} k - \omega \right ) } \frac{(p^1)^2-(p^2)^2}{2 \Bar{p}_0} \frac{\partial f^{(0)}}{\partial p^3} \Tilde{h}_{11} \\
& + 2 \frac{ck}{ \left ( \frac{c \, p^3}{\Bar{p}_0} k - \omega \right ) } \frac{p^1p^2}{\Bar{p}_0} \frac{\partial f^{(0)}}{\partial p^3} \Tilde{h}_{12}
\,. \label{e243}
\end{split}
\end{equation}
At first sight, it could seems that, when substituted in equations (232) and (233), equation (243) would leave to a linear homogeneous system for $\Tilde{h}_{11}$ and $\Tilde{h}_{12}$. However, assuming an even zero order distribution function satisfying $\partial f^{(0)} / \partial p^i \sim p^i$ (the case of Maxwell and Synge-Jüttner distributions, for example), many of the several resulting momentum space integrals vanishes by virtue of parity, and the only non vanishing resulting integrals are given by
\begin{equation}
\begin{split}
\mathcal{A} 
= \int \frac{ \left [ (p^1)^2-(p^2)^2 \right ]^2}{2 (\Bar{p}_0)^3} f^{(0)} \, d^3p
\, \label{e244}
\end{split}
\end{equation}
\begin{equation}
\begin{split}
\mathcal{A}' 
= \int \frac{ (p^1)^2-(p^2)^2}{\Bar{p}_0} \left(p^1 \frac{\partial}{\partial p^1} - p^2 \frac{\partial}{\partial p^2} \right)  f^{(0)} \, d^3p
\, \label{e245}
\end{split}
\end{equation}

\begin{equation}
\begin{split}
\mathcal{B} 
= \int \frac{2 (p^1)^2(p^2)^2 }{(\Bar{p}_0)^3} f^{(0)} \, d^3p
\, \label{e246}
\end{split}
\end{equation}
\begin{equation}
\begin{split}
\mathcal{B}' 
= \int \frac{2 p^1p^2 }{\Bar{p}_0} \left(p^2 \frac{\partial}{\partial p^1} + p^1 \frac{\partial}{\partial p^2} \right)  f^{(0)} \, d^3p
\, \label{e247}
\end{split}
\end{equation}
\begin{equation}
\begin{split}
\mathcal{R} 
= c\int \frac{1}{u-\omega / k} \frac{ \left [ (p^1)^2-(p^2)^2 \right]^2}{2 (\Bar{p}_0)^2} \frac{\partial f^{(0)}}{\partial p^3}  \, d^3p
\, \label{e248}
\end{split}
\end{equation}
\begin{equation}
\begin{split}
\mathcal{S} 
= c\int \frac{1}{u- \omega / k } \frac{2 (p^1)^2(p^2)^2 }{(\Bar{p}_0)^2} \frac{\partial f^{(0)}}{\partial p^3}  \, d^3p
\,, \label{e249}
\end{split}
\end{equation}
with $u = cp^3 / \Bar{p}_0$, the $z$ component of the 3-velocity. In terms of the above integrals, equations (232) and (233) gives the two \textit{independent} dispersion relations for  $\Tilde{h}_{11}$ and $\Tilde{h}_{12}$, respectively:
\begin{equation}
\begin{split}
\omega^2 - c^2k^2 = \frac{16\pi G n_0}{c} \left ( \mathcal{A} - \mathcal{A}' + \mathcal{R} \right ) \, \label{e250}
\end{split}
\end{equation}
and
\begin{equation}
\begin{split}
\omega^2 - c^2k^2 = \frac{16\pi G n_0}{c} \left ( \mathcal{B} - \mathcal{B}' + \mathcal{S} \right ) \,. \label{e251}
\end{split}
\end{equation}
Indeed, there is no system of equations to be solved here. Shortly, we will deal with the integrals (244)-(249) and show that the dispersion relations (250) and (251) are exactly the same, as one could expect. For now, it is important to observe that the integrals $\mathcal{A}$ , $\mathcal{A}'$, $\mathcal{B}$ and $\mathcal{B}'$ results in just real functions of the physical parameters of the plasma (as the temperature and the electron mass), while $\mathcal{R}$ and $\mathcal{S}$ contains the denominator $u - \omega / k$, causing the function do be integrated to become singular for $u = \omega / k$, the phase velocity of the wave (as in the non relativistic theory).  This singularity, as is well known, is related to the Landau damping. In the following subsection we will solve the integrals found in the limit of low temperatures and, with this, we will investigate the possibility of the Landau damping in the electron-positron plasma.

\subsection{Evaluation of the integrals. On the Landau damping}

We now proceed to approximately evaluate the 
integrals (244)-(248) and to investigate the dispersion relations (250) and (251). First, to show that the two dispersion relations are indeed the same, we adopt a spherical coordinate system for $\mathbf{p}$, with the wave vector $\mathbf{k}$ oriented along the $z$ axis. Furthermore, to perform more concrete calculations, when appropriate, we will assume the Synge-Jüttner zero order distribution function
\begin{equation}
\begin{split}
f^{(0)}_{SJ}(\mathbf{p}) = \frac{1}{4 \pi \, m^3c^3} \frac{\mu}{K_2(\mu)} e^{-\mu \gamma}
\,, \label{e252}
\end{split}
\end{equation}
where $\mu = mc^2 / k_B T$ is the temperature parameter and $K_2 (\mu)$ is the modified Bessel function of second kind, of order 2. Observe that $\gamma (p) = \sqrt{1 + p^2 / m^2c^2}$ is the usual Lorentz factor (not to be confused with the gamma defined in equation (155)), with $\mathbf{p} = \gamma m \mathbf{v}$, and $\mathbf{v}$ is the 3-velocity. With the mentioned choice of coordinates we have
\begin{equation}
\begin{split}
p^1 = p_1 = p_x = p \, sin \theta \, cos \phi
\,, \label{e253}
\end{split}
\end{equation}
\begin{equation}
\begin{split}
p^2 = p_2 = p_y = p \, sin \theta \, sin \phi
\,, \label{e254}
\end{split}
\end{equation}
\begin{equation}
\begin{split}
p^3 = p_3 = p_z = p \, cos \theta
\,, \label{e255}
\end{split}
\end{equation}
where $\theta$ and $\phi$ are the polar and azimuthal angles, respectively. Thus, from equations (244)-(248) with $\Bar{p}_0 = \gamma mc$ we get, after some algebraic manipulations and integration in the azimuthal angle,
\begin{equation}
\begin{split}
\mathcal{A} = \mathcal{B} = \frac{\pi}{2 \, m^3 c^3} \int_0^\infty dp \, \frac{p^6 f^{(0)}}{\gamma^3} \int_0^\pi d\theta \, sin^5 \theta \, \label{e256}
\end{split}
\end{equation}
\begin{equation}
\begin{split}
\mathcal{A}' = \mathcal{B}' = - \frac{\pi \, \mu}{m^3 c^3} \int_0^\infty dp \, \frac{p^6 f^{(0)}}{\gamma^2} \int_0^\pi d\theta \, sin^5 \theta \, \label{e257}
\end{split}
\end{equation}
\begin{equation}
\begin{split}
\mathcal{R} = \mathcal{S} = \frac{\pi \, c}{2 \, m^2 c^2} \int_0^\infty dp \, \frac{p^6}{\gamma^2} \int_0^\pi d\theta \, \frac{ sin^5 \theta}{u - \omega / k } \frac{\partial f^{(0)}}{\partial p_z} \, \label{e258}
\end{split}
\end{equation}
with $u = p_z / \gamma m$. The result (257) was obtained employing the relativistic distribution (252), but remains valid in the non-relativistic limit with $\gamma \rightarrow 1$, for which we can use the Maxwell distribution. With the results (256)-(258), and taking into account equations (250) and (251), it is clear that, in fact, the two polarization states of gravitational waves obeys the same dispersion relation, as expected.  

To analyse the behavior of the dispersion relation (250), it is instructive do perform the integrals (256)-(258) assuming low particle speeds, that is, a low temperature plasma so that $k_BT << mc^2$ (and $\mu >> 1$). In this non-relativistic limit it can be shown that the Synge-Jüttner function becomes the Maxwell distribution
\begin{equation}
\begin{split}
f^{(0)}_M(\mathbf{p}) = \frac{1}{(2 \pi \, m \, k_B T)^{3/2}} e^{-p^2/2m \,k_B T}
\,. \label{e259}
\end{split}
\end{equation}
With the distribution (259) and $\gamma = 1$, the integrals (256) and (257) can be solved quickly by elementary methods and employing the tabulated integral
\begin{equation}
\begin{split}
\int_0^\infty x^{2n}e^{-x^2 / a} dx = \frac{(2n - 1)!! \sqrt{\pi}}{2^{(n+1)}} a^{(2n + 1)/2}
\, \label{e260}
\end{split}
\end{equation}
for $n$ even. The integration given by equation (258) is more tractable in its original rectangular forms (equations (248) and (249)) as the Maxwell distribution can be factorized in the form
\begin{equation}
\begin{split}
f^{(0)}_M(\mathbf{p}) = f^{(0)}_x(p_x) f^{(0)}_y(p_y)
f^{(0)}_z(p_z)
\,, \label{e261}
\end{split}
\end{equation}
with
\begin{equation}
\begin{split}
f^{(0)}_i(p_i) = \frac{1}{(2 \pi \, m \, k_B T)^{1/2}} e^{-p_i^2/2m \,k_B T}
\,. \label{e262}
\end{split}
\end{equation}
The results are
\begin{equation}
\begin{split}
\mathcal{A} = \frac{2 \, (k_B T)^2}{m c^3} \, \label{e263}
\end{split}
\end{equation}
\begin{equation}
\begin{split}
\mathcal{A}' = - \frac{4 \mu \, (k_B T)^2}{m c^3} \, \label{e264}
\end{split}
\end{equation}
\begin{equation}
\begin{split}
\mathcal{R} = \frac{2 \, (k_B T)^2}{mc} \int_{-\infty}^\infty \frac{dF(u)/du}{u - \omega/k}du \,. \label{e265}
\end{split}
\end{equation}
In equation (265) was used the distribution function $F(u)$ for the velocity $u$, defined by
\begin{equation}
\begin{split}
F(u) \, du = f^{(0)}(p_z) \, dp_z
 \,. \label{e266}
\end{split}
\end{equation}
With this, as is well known, we have
\begin{equation}
\begin{split}
F(u) = mf^{(0)}(mu) = \left (\frac{m}{2 \pi \, k_B T} \right )^{1/2} e^{-mu^2/2 \,k_B T} \,. \label{e267}
\end{split}
\end{equation}
We are now in place to return to the distribution function and discuss the issue of the non-collisional damping of gravitational waves for a low temperature electron-positron plasma. Substituting the equations (263)-(265) in equation (250), we find
\begin{equation}
\begin{split}
& \omega^2 - c^2k^2  = 2\, \omega_g^2 \left( \frac{2}{\mu} + \frac{1}{\mu^2} \right ) \\ + \frac{2 \, \omega_g^2 c^2} {\mu^2} &
\left ( \mathcal{P} \int_{-\infty}^\infty \frac{dF/du}{u 
- \omega/k}du + i \pi 
        \begin{matrix}
        \frac{dF}{du}\vline
        \end{matrix}
\, _{u \, = \, \omega / k} \right )
\,, \label{e268}
\end{split}
\end{equation}
where we used the usual prescription
\begin{equation}
\begin{split}
\int_{-\infty}^\infty \frac{F(z)}{z 
- z_0}dz = \mathcal{P} \int_{-\infty}^\infty \frac{F(z)}{z 
- z_0}dz + i \pi F(z_0) 
\,, \label{e269}
\end{split}
\end{equation}
in which $\mathcal{P}$ stands for Cauchy principal value. Furthermore, we define the gravitational plasma frequency $\omega_g$ as
\begin{equation}
\begin{split}
\omega_g^2 = 16 \pi G\,m n_0
\,. \label{e270}
\end{split}
\end{equation}
Compare the definition above with that given in references \cite{serv2} and \cite{inan2}, and observe that our definition of $\omega_g$ would almost be obtained in the form (270) by taking the electron plasma frequency given by $\omega_p^2 = e^2 n_0 / m \, \epsilon_0$ and making some formal substitutions based on the comparison between the Newton universal gravitation law and Coulomb law, namely $\epsilon_0 \rightarrow 1/4\pi G$ and $e^2 \rightarrow m^2$.

Now, assuming $(\omega / k)^2 >> k_B T / m$, it is known that \cite{chen,bit}
\begin{equation}
\begin{split}
\mathcal{P} \int_{-\infty}^\infty \frac{dF/du}{u 
- \omega/k}du = \frac{k^2}{\omega^2} + \frac{3}{2} \frac{k^4}{\omega^4} \frac{c^2}{\mu} + \, ...
\,. \label{e271}
\end{split}
\end{equation}
Inserting the series expansion (271) in equation (268) and retaining only terms proportional to $1/\mu$ and $1/\mu^2$, we are led to
\begin{equation}
\begin{split}
\omega^2 - c^2k^2  = 
& \frac{4\,\omega_g^2}{\mu} +\frac{2\,\omega_g^2}{\mu^2} \frac{\left (\omega^2 + c^2 k^2 \right )}{\omega^2}
\\ + 
& \frac{2i\pi\,\omega_g^2c^2}{\mu^2}
        \begin{matrix}
        \frac{dF}{du}\vline
        \end{matrix}
\, _{u \, = \, \omega / k}
\,. \label{e272}
\end{split}
\end{equation}
Following the usual procedure, we now write $\omega = \omega_R + i\omega_I$, with $\omega_R$ and $\omega_I$ respectively the real and imaginary parts of $\omega$, and assume $\omega_I << \omega_R$. With $\omega$ substituted in equation (272) we arrive at
\begin{equation}
\begin{split}
\omega_R^2 - c^2k^2  = \frac{4\,\omega_g^2}{\mu}
\, \label{e273}
\end{split}
\end{equation}
and
\begin{equation}
\begin{split}
\omega_I = 
\frac{\pi\,\omega_g^2c^2}{\mu^2 \omega_R}
        \begin{matrix}
        \frac{dF}{du}\vline
        \end{matrix}
\, _{u \, = \, \omega_R / k}
\, \label{e274}
\end{split}
\end{equation}
by discarding terms proportional to $1/\mu^2$ and to any power of $\omega_I$ in the expression for $\omega_R$, and retaining only the leading terms to determine $\omega_I$. To find the damping parameter $\omega_I$, we must first find the expression for the phase velocity $v_\phi = \omega_R/k$ of the gravitational waves from equation (272), and then calculate the derivative in equation (274) applied at $v_\phi$. It happens, however, that equation (273) leads to a phase velocity greater than $c$ (corresponding to a refractive index less than unity):
\begin{equation}
\begin{split}
v_\phi = c \left ( 1 + \frac{4\,\omega_g^2}{\mu c^2 k^2} \right )^{1/2} > c 
\,. \label{e275}
\end{split}
\end{equation}
Thus, as the relativistic particle dynamics does not allows $u>c$ and no physical distribution function can really extend to this superluminal regime (although Maxwell's does, since it is a non-relativistic distribution), we conclude that $\omega_I = 0$, and there is no Landau damping for gravitational waves. Physically, it just means that, as electrons and positrons are not allowed to travel in the direction of propagation of the wave with the same speed as it, the resonant wave-particle coupling can not occur, and no energy exchange between particles and waves can take place (the same as for electromagnetic waves). 

Let's discuss a little deeper this issue. Despite the fact that for the system we are studying there is no Landau damping, let's pretend for a moment that $\omega_I \neq 0$. From equation (109) (see Section 2, part D) we can readily establishes the following expression for the gravitational instantaneous energy exchange rate (in watt per cubic meter), in the weak field approximation:
\begin{equation}
\begin{split}
\Gamma_g = \frac{c^4}{16\pi G} \frac{\partial h_{\mu \nu}}{\partial t}
G^{\mu \nu} 
\,. \label{e276}
\end{split}
\end{equation}
It is the density of work realized by (or on) the gravitational waves, that is, the rate of gravitational energy loss (or gain) per unit volume. Now, employing equation (47) for the spatial components of the Einstein tensor, we find for gravitational waves
\begin{equation}
\begin{split}
\Gamma_g = \frac{c^4}{32\pi G} \frac{\partial h_{ij}}{\partial t}
\Box h^{ij} 
\,. \label{e277}
\end{split}
\end{equation}
As usual when dealing with wave phenomena, we are just interested in the mean value of $\Gamma_g$ over a period. To achieve this value, we can proceed in two ways. The first, is to take the real part of the complex plane wave (192), that could be
\begin{equation}
\begin{split}
h_{ij} = \epsilon_{ij} e^{\omega_I t} cos(kz - \omega_Rt) 
\, \label{e278}
\end{split}
\end{equation}
if we assume real $\epsilon_{ij}$, substitute their relevant derivatives in equation (277) and take the mean values of the trigonometric functions that appear in the calculations. The second approach is a bit more economic and clean, for it does not requires any explicit calculation of mean values. In this method, we persist in writing $h_{ij}$ in the complex form (192), and apply the prescription \cite{jack}
\begin{equation}
\begin{split}
\left < \frac{\partial h_{ij,\text{Real}}}{\partial t}
\Box h^{ij}_\text{Real} \right > = \frac{1}{2} \mathrm{Re} \left ( \frac{\partial h^*_{ij}}{\partial t}
\Box h^{ij} \right )
\,. \label{e279}
\end{split}
\end{equation}
where the symbol $<>$ represents the mean value over a period. With formula (279) inserted in equation (277) we find
\begin{equation}
\begin{split}
\left <\Gamma_g \right > = \frac{c^4}{64\pi G} \,\mathrm{Re} \left ( \frac{\partial h^*_{ij}}{\partial t}
\Box h^{ij} \right )
\,. \label{e280}
\end{split}
\end{equation}
Whatever the method used, the result obtained is
\begin{equation}
\begin{split}
\left < \Gamma_g \right > =  \frac{\omega_I c^2}{32\pi G} \left ( \omega_R^2 - \frac{2\omega_g^2}{\mu}\right )  \, \epsilon_{ij} \epsilon^{ij} e^{2\,\omega_It}
\,, \label{e281}
\end{split}
\end{equation}
where only leading terms were retained and the dispersion relation (273) was used. Note that, as the smallest possible value of  $\omega_R^2$ is $4 \, \omega_g^2/\mu$, the bracket in equation (281) is always positive, and the sign of $\left <\Gamma_g\right>$ is linked to the sign of $\omega_I$. If  $\omega_I < 0$, the gravitational wave is damped and $\left <\Gamma_g\right> < 0$, indicating a decrease of the gravitational energy density with time. If, on the other hand,  $\omega_I > 0$, the gravitational wave would presents an instability and $\left <\Gamma_g\right> > 0$, indicating an increase of the gravitational energy density with time. Of course, as the correct value for the Landau parameter is $\omega_I = 0$, then $\left <\Gamma_g\right> = 0$, agreeing with the previous affirmation about the non-existence of energy exchange between the gravitational wave and the particles of the plasma. Just to mention, observe that the opposite limit, with $(\omega / k)^2 << k_B T / m$, can not be described properly employing the Maxwell distribution, for it would requires particle speeds typically greater than the phase velocity of the wave which, in turn, is greater than $c$.

The physical conclusions found so far are interesting and reasonable, but applies only to a low temperature plasma in the limit $(\omega / k)^2 >> k_B T / m$. A general, fully relativistic treatment valid for a wide range of temperatures and frequencies, requires to solve the integrals (256)-(258) using the Synge-Jüttner distribution (252). These calculations, although difficult, can potentially reveal a richer structure for the dispersion relation, and will be presented and discussed in a future opportunity. 
\\
\section{Conclusion}

The linear regime of Einstein field equations and the problem of gauge invariance of the underlying theory were revised and meticulously analysed employing the Helmholtz decomposition scheme for vectors and second order tensors. In this regime, the field equations are splited up in a set of differential equations for two scalar, one vector and one tensor gauge invariant gravitational potentials, the first three obeying Poisson-type equations and the last satisfying a non-homogeneous wave equation, being associated to gravitational radiation. Although the metric is dependent on a choice of gauge (as required by the principle of equivalence), the Einstein tensor is clearly a gauge invariant object, being physical significant. The problem of gravitational waves propagating in free space were revised under the gaze of the gauge invariant theory, showing that this methodology it very simple and physically illuminating, much better than be worried about choosing this or that gauge and embarrassed for differentiating real gravitational wave effects from spurious coordinate choice ones. From the theory we can quickly obtain the relevant field equation for a given system or problem. We set a general equation describing gravitational energy exchanges and some useful formulae to write down the source terms for gravitation employing Fourier transforms. Furthermore, we set some general expressions for the Christoffel symbols and for volume elements in terms of gauge invariant potential, necessary to correctly develop the relevant Einstein, Vlasov and Maxwell equations and evaluate momentum space integrals.

After briefly revising the Einstein-Vlasov-Maxwell system in the linear regime, we apply the theory for describing electrostatic and gravitational waves in an electron-positron plasma, showing that, in this case, to the first order, there is a complete decoupling between electric and gravitational oscillation. For low temperatures, we find an dispersion relation for the gravitational waves assuming $(\omega / k)^2 >> k_B T / m$, showing that these waves are not damped, and so have their energy conserved. The momentum space integrals involving the Synge-Jüttner distribution were not solved exactly, a job that is left to a future work and promises to reveal a structurally richer dispersion relation.

As mentioned in the introduction, our aim was to consistently bring together plasma kinetic theory and the linear theory of gravitation in terms of gauge invariant potentials, establishing a general, simple and secure methodology to deal in equal footing with oscillations of any (radiative or not) components of the gravitational field. In future works, the methodology presented here will be applied to several problems with isotropic and non-isotropic plasmas, magnetized or not, involving the whole set of gravitational gauge invariant potentials. We hope that the present paper will help to clarify, to anyone interested in the subject, some issues involving the gauge invariance in general relativity, motivating some plasma physicist to take a tour and make some adventures in general relativity and general relativistic physicist to make some trials in plasma physics. We hope too that our systematic and (more or less) complete presentation could facilitates every physicist interested in general relativistic plasma and kinetic theory to pursue their goals, and motivates futures works, in the way the theory discussed here be extended to the quantum level.

Finally, it is worth mentioning that there are several alternative formulations of the theory of general relativity, such as the ADM formalism. In this version, the covariant general relativistic plasma equations can be cast into more familiar special relativistic forms. As a promising avenue, it can be investigated more deeply, on how the theory fits in this (or other) alternative formulation and how to employ it in order to explore gravitational field and plasma dynamics in terms of gauge invariant quantities. Unfortunately, we have not yet been able to make efforts in this direction and, while we can apologize for this, we also extend an invitation to other researchers to get involved with this subject.

\acknowledgments

LB acknowledges CAPES (Coordenação de Aperfeiçoamento de Pessoal de Nível Superior) for financial support. 
FH acknowledges CNPq (Conselho Nacional de Desenvolvimento Cient\'{\i}fico e
Tecnol\'ogico) for financial support.

\end{document}